\documentclass[fleqn,usenatbib]{mnras}

\usepackage{newtxtext,newtxmath}

\usepackage[T1]{fontenc}

\DeclareRobustCommand{\VAN}[3]{#2}
\let\VANthebibliography\thebibliography
\def\thebibliography{\DeclareRobustCommand{\VAN}[3]{##3}\VANthebibliography}

\usepackage{graphicx}	
\usepackage{amsmath}	
\usepackage{orcidlink}
\usepackage{subfiles}
\usepackage{subcaption}
\usepackage{multirow}
\usepackage{booktabs}

\title[Free-Form Analysis of Jackpot's Subhalo]{Probing Dark Matter Substructures with Free-Form Modelling: \\A Case Study of the `Jackpot' Strong Lens}
\author[Cao et al.]{Xiaoyue Cao\orcidlink{0000-0003-4988-9296}$^{1,2,3}$\thanks{E-mail: xycao@zzu.edu.cn, xycao@nao.cas.cn}, 
Ran Li\orcidlink{0000-0003-3899-0612}$^{4,3}$\thanks{E-mail: liran@bnu.edu.cn},
James W. Nightingale$^{5,6,7}$\orcidlink{0000-0002-8987-7401},
Richard Massey$^{6,7}$\orcidlink{0000-0002-6085-3780}, 
Qiuhan He$^{6}$\orcidlink{0000-0003-3672-9365}, \newauthor
Aristeidis Amvrosiadis$^{6}$\orcidlink{0000-0002-4465-1564}, 
Andrew Robertson$^{8}$\orcidlink{0000-0002-0086-0524},
Shaun Cole$^{6}$\orcidlink{0000-0002-5954-7903},
Carlos S. Frenk$^{6}$\orcidlink{0000-0002-2338-716X},
Xianghao Ma$^{2,3,4}$, \newauthor
Leo W.H. Fung$^{6,7}$\orcidlink{0000-0002-5899-3936},
Maximilian von Wietersheim-Kramsta$^{6,7}$\orcidlink{0000-0003-4986-5091},
Samuel C. Lange$^{6}$\orcidlink{0009-0007-0679-818X},
Kaihao Wang$^{2,3,4}$,\newauthor
Liang Gao$^{4,1,3}$
\\
$^{1}$Institute for Astrophysics, School of Physics, Zhengzhou University, Zhengzhou, 450001, China\\
$^{2}$School of Astronomy and Space Science, University of Chinese Academy of Sciences, Beijing 100049, China\\
$^{3}$National Astronomical Observatories, Chinese Academy of Sciences, 20A Datun Road, Chaoyang District, Beijing 100012, China\\
$^{4}$ School of Physics and Astronomy, Beijing Normal University,  Beijing 100875, China\\
$^{5}$ School of Mathematics, Statistics and Physics, Newcastle University, Newcastle upon Tyne, NE1 7RU, UK \\
$^{6}$ Institute for Computational Cosmology, Department of Physics, Durham University, South Road, Durham DH1 3LE, UK\\
$^{7}$ Centre for Extragalactic Astronomy, Department of Physics, Durham University, South Road, Durham DH1 3LE, UK \\
$^{8}$ Observatories, Carnegie Institution for Science, 813 Santa Barbara Street, Pasadena, CA 91101, USA \\
}

\date{Accepted XXX. Received YYY; in original form ZZZ}

\pubyear{\the\year{}}

\begin{document}
\label{firstpage}
\pagerange{\pageref{firstpage}--\pageref{lastpage}}
\maketitle

\begin{abstract}
Characterising the population and internal structure of sub-galactic halos is critical for constraining the nature of dark matter. These halos can be detected near galaxies that act as strong gravitational lenses with extended arcs, as they perturb the shapes of the arcs. However, this method is subject to false-positive detections and systematic uncertainties, particularly degeneracies between an individual halo and larger-scale asymmetries in the distribution of lens mass. We present a new free-form lens modelling code, developed within the framework of the open-source software \texttt{PyAutoLens}, to address these challenges. Our method models mass perturbations that cannot be captured by parametric models as pixelized potential corrections and suppresses unphysical solutions via a Mat\'ern regularisation scheme that is inspired by Gaussian process regression. This approach enables the recovery of diverse mass perturbations, including subhalos, line-of-sight halos, external shear, and multipole components that represent the complex angular mass distribution of the lens galaxy, such as boxiness/diskiness. Additionally, our fully Bayesian framework objectively infers hyperparameters associated with the regularisation of pixelized sources and potential corrections, eliminating the need for manual fine-tuning. By applying our code to the well-known `Jackpot' lens system, SLACS0946+1006, we robustly detect a highly concentrated subhalo that challenges the standard cold dark matter model. This study represents the first attempt to independently reveal the mass distribution of a subhalo using a fully free-form approach.
\end{abstract}




\section{Introduction}
\label{sec:intro}
The standard cosmological model has achieved remarkable success \citep{Planck2018}, as it explains numerous observations, such as the rotation curves of disk galaxies \citep{Rubin1991}, the clustering properties of galaxies \citep{Springel2005}, the properties of the hot gas in galaxy clusters \citep{Ettori2013}, and the cosmic microwave background \citep{Planck2013}. The standard cosmological model assumes that approximately 85 percent of the matter in the universe is the so-called "dark matter" \citep{Planck2018}. Dark matter neither emits light nor interacts electromagnetically with baryonic matter. Instead, its presence is inferred solely from gravitational effects \citep{Arbey2021}. The fundamental nature of dark matter remains elusive; it may be primordial black holes \citep{Carr2016} or some undiscovered particles beyond the standard model of particle physics \citep{Feng2010}. Since different types of dark matter produce distinct cosmological structures, such as dark matter halos and subhalos, astronomical observations can serve as an effective means to constrain the properties of dark matter \citep{Frenk2012}. The popular probes currently used to infer the properties of dark matter include stellar streams in the local group \citep{Ibata2001, Bonaca2019}, the Lyman-alpha forest \citep{Seljak2006}, the luminosity function of Milky Way satellite galaxies \citep{Nadler2021, Newton2021}, and strong gravitational lensing \citep{Vegetti2023}. Among these, the strong gravitational lensing technique offers a distinct advantage, as it can infer the properties of dark matter solely based on the gravitational distortion effects of dark matter structures, without relying on complex baryonic physics, and it can be applied to higher redshifts.

Strong gravitational lensing occurs when light from a background source is significantly distorted by a massive foreground deflector, such as a galaxy or a galaxy cluster, forming multiple images or extended arcs \citep{Schneider1992_textbook, LensingBook, Meneghetti22_textbook}. Smaller dark matter halos within the main deflector (subhalos) or along the line of sight \citep[``LOS-halos'', ][]{Li2017, Despali2018_los, He2022_los} can further perturb the lensing signal, creating observable anomalies in lensed images, time delays, and flux ratios between multiple images. These anomalies serve as valuable probes for constraining the properties of dark matter \citep{Vegetti2009_stat, Li2016, He2022_abc, Keeton2009, Liao2018, Mao1998, Birrer2017_abc, Gilman2018, Hsueh2020}. Current time-delay measurements, typically with uncertainties on the order of a few days \citep{Millon2020_td_ms}, lack the precision required to detect subhalo-induced signals, which are often on the order of fractions of a day. Flux ratio anomalies, theoretically most sensitive to subhalo perturbations due to their direct dependence on the second derivative of the lensing potential, generally cannot precisely constrain the properties (e.g., mass and position) of individual subhalos \citep[e.g.][]{Nierenberg2014}. This limitation arises from the sparsity of observational data (usually restricted to positions and fluxes of lensed images), which does not provide sufficient constraints to resolve degeneracies either among subhalo internal properties (e.g., mass-position degeneracy) or between the subhalo and the main lens galaxy \citep{Fadely2012, Hsueh2020, Cohen2024}. Consequently, subhalo properties inferred from flux ratio anomalies are typically derived statistically—finding multiple feasible solutions across a range of subhalo masses and positions—instead of precise individual measurements \citep{Dalal2002, Hsueh2020, Gilman2020}. In contrast, the brightness distribution along distorted lensed arcs, typically comprising several thousand pixels, can effectively break the degeneracies encountered by flux ratio anomalies. Thus, analysing extended arcs offers a more reliable method for precisely measuring the properties of individual subhalos through strong gravitational lensing \citep[e.g.,][]{Vegetti2010, Vegetti2012, Hezaveh2016_alma_subhalo, Ritondale2019, Nightingale2024}.

Subhalo detections using distorted lensed arcs have been widely employed to constrain dark matter properties \citep[e.g.,][]{Vegetti2014, Vegetti2018, Enzi2021}. Since the distortions induced by small-mass subhalos are subtle, sophisticated lens modelling methods are essential for extracting these weak signals. Currently, two primary approaches are employed: parametric and free-form methods. The parametric method uses explicit parametric models to describe the mass distribution of both the main lens galaxy and the presumed subhalo. In this approach, a subhalo is considered detected if a model including a subhalo provides a significantly better fit to the observational data, as measured by an increase in Bayesian evidence, compared to a model without a subhalo \citep{Vegetti2009_method, Ritondale2019, Nightingale2024}. In contrast, the free-form method approximates the mass distribution of the main lens galaxy using a smooth parametric model while treating perturbative signals not captured by the parametric model as pixelized linear corrections to the lensing potential\footnote{This free-form approach to the pixelized perturbative lensing potential is widely known as the potential correction method. Other types of free-form approaches have also been employed in strong lensing analyses, which represent the mass (instead of the lensing potential) on a grid of pixels \citep[e.g.][]{Saha_1997} or as a linear combination of basis functions \citep[e.g.][]{Liesenborgs2006, Diego2007}.} \citep{Koopmans2005, Suyu2006_pt, Suyu2009_pt, Vegetti2009_method, Ritondale2019, Vernardos2022, Galan2022_wavelet, LenCharm}. Subhalos appear as localised positive mass clumps in the resulting mass perturbation map. While the parametric method is effective at quantifying the properties of subhalo perturbers, it is prone to false positives, where a subhalo is incorrectly inferred to exist \citep{Ritondale2019}. This issue stems from systematic errors in the parametric model, particularly degeneracies between the mass distribution of the main lens galaxy and the subhalo. If the main lens galaxy's mass model is oversimplified, a false-positive subhalo signal may arise to compensate for missing complexity in the lens galaxy \citep{He2023_cosmos_sim, Nightingale2024, Riordan2024}. Although adding complexity to the lens mass model can test the robustness of subhalo detections \citep{Lange2024}, it is often unclear whether the added complexity is sufficient to eliminate all false positives. Consequently, subhalo detection claims based solely on parametric methods may be met with scepticism. To address this concern, \cite{Vegetti2014} advocated for free-form approaches to validate subhalo detections obtained using parametric methods.

Regularisation, or smoothing, is essential for obtaining physically meaningful solutions in free-form approaches \citep{Wallington1996, Warren2003}. The classical potential correction method begins with an initial estimate of the main lens mass distribution and the source brightness distribution. A strong regularisation strength is empirically applied to the pixelized lensing potential perturbations, producing a minor correction to the lensing potential \citep{Koopmans2005, Vegetti2009_method}. This correction is subsequently used to refine both the lens mass and source light models, initiating a new iteration of potential correction. The process is repeated iteratively until convergence is achieved, defined as the point where additional corrections no longer result in significant improvements to the model fit \citep{Koopmans2005}. A major limitation of this iterative algorithm is its vulnerability to instability: artefacts caused by an inappropriate choice of regularisation strength can accumulate over iterations, ultimately compromising the modelling process. \citet{Vernardos2022} introduced a framework to objectively determine the regularisation strength for pixelized lensing potential perturbations by maximising Bayesian evidence and directly solving for the potential correction, thereby avoiding the need for unstable iterative algorithms. However, the Gaussian and Exponential regularisation techniques used in their study are more effective for extended perturbations but perform poorly when applied to localised features, such as those caused by subhalos\footnote{Further demonstration of this limitation is provided in Section~\ref{sec:res_inv_dpsi_only}.}.

The Mat\'ern kernel is a widely used covariance function that imposes smoothness constraints in free-form reconstruction via Gaussian process regression. This Mat\'ern covariance kernel, or regularisation, has recently been applied to pixelised source reconstruction in strong gravitational lensing analyses \citep{Galan_2024, Enzi2024, LenCharm}. Whilst \citet{Vernardos2022} primarily investigated the advantages of Gaussian and exponential forms of regularisation---special cases of the more general Mat\'ern kernel---for reconstructing pixelised sources and mass models, they also proposed that the Mat\'ern kernel merits further exploration. In this work, we present a new free-form lens modelling code built upon the open-source software \texttt{PyAutoLens} \citep{Nightingale2015, PyAutoLens_1, PyAutoLens_2}. By employing a generalised Mat\'ern regularisation for the pixelised mass model (in the form of potential corrections) \citep{Rasmussen2006_gpr} and determining the regularisation parameters objectively through the Bayesian evidence, our code automatically recovers diverse forms of mass perturbation signals without requiring manual fine-tuning. To validate our method, we first test it on mock lensing systems, demonstrating its robustness in recovering various forms of lensing potential perturbations commonly encountered in strong lensing studies. We then apply our method to the strong lens system SLACS0946+1006 \citep{Gavazzi2008, Smith2021}, also known as the `Jackpot' lens, and characterise the mass perturbations in this system using a fully free-form approach.

This paper is organised as follows. In Section~\ref{sec:method}, we describe the methodology for the potential correction. Section~\ref{sec:dataset} describes the datasets employed in this study, including mock datasets and the real ``Jackpot'' lens system (SLACS0946+1006). In Section~\ref{sec:result}, we first evaluate the performance of our method in quantitatively recovering mass perturbation signals using mock datasets and then present a free-form analysis of the real lens system SLACS0946+1006. The strengths, limitations, and future prospects of our method are discussed in Section~\ref{sec:discuss}, and a concise summary is provided in Section~\ref{sec:summary}. Unless otherwise stated, we adopt the posterior median as the default point estimate for the model parameters. Throughout this work, we assume a flat $\Lambda$CDM cosmology with $\Omega_{\rm m} = 0.3$, $\Omega_\Lambda = 0.7$, and $H_0 = 70\, \mathrm{km}\,\mathrm{s}^{-1}\,\mathrm{Mpc}^{-1}$. All codes and relevant data used in this work are publicly available at the following link: \url{https://github.com/caoxiaoyue/lensing_potential_correction}.
\section{Methodology}
\label{sec:method}
This section provides a concise review of the framework for the potential correction method. Section~\ref{sec:linear_approx} presents the linear approximation that underpins the potential correction method. Section~\ref{sec:inv_dpsi_only} outlines the framework for only solving lensing potential perturbations given an initial estimate of the lens mass and source. Section~\ref{sec:inv_dpsi_src} describes how to simultaneously solve for source and lensing potential perturbations, accounting for their covariance. The configuration of the free-form model, including the pixelization and regularisation strategies, is presented in Section~\ref{sec:reg_func}.

\subsection{Linear Approximation Relation}
\label{sec:linear_approx}
In galaxy-galaxy strong lens modelling, a smooth parametric mass model for the lens galaxy, such as the power-law plus shear model, typically provides a satisfactory fit to the global morphological features of the lensing image \citep[e.g.][]{Shajib2021, Etherington2022, Tan2024}. Residuals between the model and data images are usually small, suggesting that any omitted mass complexities unaccounted for in this smooth lens model (commonly referred to as the macro model) are minor and can be treated as small perturbations to the lensing potential. The potential correction method aims to identify a set of such perturbations defined on a pixelized grid, reducing image residuals to a level consistent with observational noise \citep{Koopmans2005}. Such perturbations can be identified using a linear approximation that relates the lensing potential perturbation, the source brightness gradient, and the image residual, as derived next.

Let $R^L(\boldsymbol{\theta})$ denote the residuals in the image produced by the macro model at the image-plane angular position $\boldsymbol{\theta}$, defined as
\begin{equation}
R^L(\boldsymbol{\theta}) \equiv I^T(\boldsymbol{\theta}) - I^M(\boldsymbol{\theta}) + N(\boldsymbol{\theta}).
\label{eq:def_res}
\end{equation}
Here, $I^T$ denotes the true brightness of the lensed images, and $I^M$ denotes the brightness predicted by the macro model. $N(\boldsymbol{\theta})$ represents the image noise realisation. The residual $R^L(\boldsymbol{\theta})$ is related to the intrinsic true source brightness distribution, $S$, via
\begin{subequations} \label{eq:res_src_rel}
\begin{align}
R^L(\boldsymbol{\theta}) - N(\boldsymbol{\theta}) &= \text{PSF} \circledast \left[S(\boldsymbol{\beta}^T) - S(\boldsymbol{\beta}^M)\right], \label{eq:res_src_rel_a} \\
&= \text{PSF} \circledast \left[\frac{\partial S}{\partial \boldsymbol{\beta}}\left(\boldsymbol{\beta}^T - \boldsymbol{\beta}^M\right)\right], \label{eq:res_src_rel_b}
\end{align}
\end{subequations}
where $\text{PSF}$ denotes the point spread function kernel, and $\circledast$ represents the convolution operator. The quantities $\boldsymbol{\beta}^T$ and $\boldsymbol{\beta}^M$ indicate the ray-traced source-plane positions corresponding to $\boldsymbol{\theta}$ under the lensing potentials given by the ground truth ($\psi^T$) and the macro model ($\psi^M$), respectively. Explicitly,
\begin{subequations} \label{eq:res_lens_map}
\begin{align}
\boldsymbol{\beta}^T &= \boldsymbol{\theta} - \frac{\partial \psi^T(\boldsymbol{\theta})}{\partial \boldsymbol{\theta}},  \label{eq:res_lens_map_a} \\
\boldsymbol{\beta}^M &= \boldsymbol{\theta} - \frac{\partial \psi^M(\boldsymbol{\theta})}{\partial \boldsymbol{\theta}}, \label{eq:res_lens_map_b} \\
\psi^T(\boldsymbol{\theta}) &\equiv \psi^M(\boldsymbol{\theta}) + \delta \psi(\boldsymbol{\theta}), \label{eq:res_lens_map_c}
\end{align}
\end{subequations}
where $\delta \psi$ denotes a small difference between the lensing potential of the macro model and the ground truth. It follows that
\begin{equation}
\boldsymbol{\beta}^T - \boldsymbol{\beta}^M = \frac{\partial \left(\psi^M - \psi^T\right)}{\partial \boldsymbol{\theta}} = -\frac{\partial \delta \psi}{\partial \boldsymbol{\theta}}.
\label{eq:src_pos_diff_dpsi}
\end{equation}
Equation~\ref{eq:res_src_rel_a} implicitly assumes that the image residual $R^L$ arises purely from the small displacements in the deflection angle caused by $\delta \psi$. This is a reasonable approximation, given that $\delta \psi$ is a small quantity. Substituting Equation~\ref{eq:src_pos_diff_dpsi} into Equation~\ref{eq:res_src_rel_b} yields
\begin{equation}
R^L(\boldsymbol{\theta}) = \text{PSF} \circledast \left[\frac{\partial S}{\partial \boldsymbol{\beta}} \cdot \left(-\frac{\partial \delta \psi}{\partial \boldsymbol{\theta}}\right)\right] + N(\boldsymbol{\theta}).
\label{eq:linear_approx}
\end{equation}
This result demonstrates that the image residuals are proportional to the product of the source brightness gradient and the gradient of the lensing potential perturbation. Note that in practical analyses, the intrinsic true source brightness distribution, $S$, is generally unknown and is typically approximated by the macro model's prediction.

\subsection{Inverting Only the Potential Perturbation}
\label{sec:inv_dpsi_only}
The primary goal of the potential correction algorithm is to solve the partial differential Equation~\ref{eq:linear_approx}, using the image residual and source gradient from the macro model. A straightforward method for solving this equation is numerical integration along the characteristic curve, as outlined in \citet{Suyu2006_pt}. However, this approach is effective only when the perturbative lensing potential is small (less than 1\%). A more robust alternative for solving equation~\ref{eq:linear_approx} is to reformulate it as the following matrix expression:
\begin{equation}
\begin{aligned}
\delta \boldsymbol{d} &= -\boldsymbol{B} \boldsymbol{D}_{\mathrm{s}} \boldsymbol{C}_\mathrm{f} \boldsymbol{D}_{\psi} \boldsymbol{\delta {\psi}} + \boldsymbol{n} \\
&= \mathbf{L_{\delta \psi}} \boldsymbol{\delta {\psi}} + \boldsymbol{n}.
\end{aligned}
\label{eq:dpsi_linear_response}
\end{equation}
We explain the meaning of each term in equation~\ref{eq:dpsi_linear_response} as follows:
\begin{enumerate}
\item{$\delta \boldsymbol{d}$ is a column vector where each element $\delta d_j$ represents the value of image residuals provided by the macro model for each image pixel, where \( j = 1, \ldots, N_d \), and \( N_d \) denotes the number of image pixels. Similarly, the vector $\boldsymbol{n}$ represents the noise realisation at each image pixel, with values drawn from the multivariate normal distribution $\mathcal{N}(0,\,\mathrm{C}_{\mathrm{D}})$, where $\mathrm{C}_{\mathrm{D}}$ is the noise covariance matrix.}
\item{$\boldsymbol{B}$ is a square matrix of dimensions $[N_d, N_d]$ constructed to account for the blurring effect of the PSF. }
\item{The vector $\delta \boldsymbol{\psi}$ comprises elements $\delta \psi_k$, each representing the value of the lensing potential perturbation at the $k$-th potential correction grid, where \( k = 1, \ldots, N_p \), and $N_p$ denotes the number of pixels in the grid. The potential correction grid is typically coarser than the native image grid, such that $N_p < N_d$, to enable a more efficient inversion process.}
\item{
$\boldsymbol{D}_{\psi}$ is a gradient operation matrix (dimensions: $2N_p \times N_p$) which, when acting on $\boldsymbol{\delta \psi}$, generates a column vector (dimensions: $2N_p \times 1$) representing the gradient of $\boldsymbol{\delta \psi}$. The generic structure of the matrix $\boldsymbol{D}_{\psi} \boldsymbol{\delta \psi}$ is shown below:
\begin{equation}
\boldsymbol{D}_{\psi} \boldsymbol{\delta \psi} = \left( \begin{array}{c}
\cdots \\
\frac{\partial \delta \psi\left(\boldsymbol{\theta_{j}^{p}}\right)}{\partial \theta_{x}} \\
\frac{\partial \delta \psi\left(\boldsymbol{\theta_{j}^{p}}\right)}{\partial \theta_{y}} \\
\frac{\partial \delta \psi\left(\boldsymbol{\theta_{j+1}^{p}}\right)}{\partial \theta_{x}} \\
\frac{\partial \delta \psi\left(\boldsymbol{\theta_{j+1}^{p}}\right)}{\partial \theta_{y}} \\
\ldots \end{array} \right).
\label{eq:def_D_dpsi}
\end{equation}
$\frac{\partial \delta \psi\left(\boldsymbol{\theta_{j}^{p}}\right)}{\partial \theta_{x}}$ and $\frac{\partial \delta \psi\left(\boldsymbol{\theta_{j}^{p}}\right)}{\partial \theta_{y}}$ represent the $x$ and $y$ components of the gradient of the potential perturbation at the $j$-th potential correction grid, denoted by $\boldsymbol{\theta_{j}^{p}}$.
}
\item{$\boldsymbol{C}_\mathrm{f}$ is an interpolation matrix with dimensions of $N_d \times N_p$ that maps a column vector defined on the coarser ``potential correction grid'' to a new vector defined at the native image resolution through linear interpolation.}
\item{$\boldsymbol{D}_{\mathrm{s}}$ is a matrix with dimensions $N_d \times 2N_d$ that stores the gradient of the source surface brightness. The generic structure of the matrix $\boldsymbol{D}_{\mathrm{s}}$ is shown below:
\begin{equation}
\boldsymbol{D}_{\mathrm{s}} = \left( \begin{array}{cccccc}
\cdots & & & & & \\
& \frac{\partial S\left(\boldsymbol{\beta_{j}^{d}}\right)}{\partial \beta_{x}} & \frac{\partial S\left(\boldsymbol{\beta_{j}^{d}}\right)}{\partial \beta_{y}} & & & \\
& & & \frac{\partial S\left(\boldsymbol{\beta_{j+1}^{d}}\right)}{\partial \beta_{x}} & \frac{\partial S\left(\boldsymbol{\beta_{j+1}^{d}}\right)}{\partial \beta_{y}} & \\
& & & & & \ldots \end{array} \right).
\label{eq:def_D_s}
\end{equation}
$\boldsymbol{\beta_{j}^{d}}$ represents the ray-traced source plane position of the $j$-th native image grid (with position denoted by $\boldsymbol{\theta_{j}^{d}}$), according to the macro model. Let $S$ be the source brightness distribution predicted by the macro model. The $x$ and $y$ components of the source's brightness gradient are given by $\frac{\partial S\left(\boldsymbol{\beta_{j}^{d}}\right)}{\partial \beta_{x}}$ and $\frac{\partial S\left(\boldsymbol{\beta_{j}^{d}}\right)}{\partial \beta_{y}}$.}
\end{enumerate}
For convenience, we abbreviate $-\boldsymbol{B} \boldsymbol{D}_{\mathrm{s}} \boldsymbol{C}_\mathrm{f} \boldsymbol{D}_{\psi}$ as $\mathbf{L_{\delta \psi}}$. The latter has dimensions of $N_d \times N_p$.

It is evident that $\delta \boldsymbol{d}$ and $\boldsymbol{\delta \psi}$ are connected via a linear mapping. Following the classical `semi-linear inversion' framework of \citet{Warren2003}, the maximum a posteriori estimate for $\boldsymbol{\delta \psi}$, for a given $\mathbf{L}_{\delta \psi}$ and regularisation $\mathbf{R}_{\delta \psi}$, can be obtained via
\begin{equation}
\boldsymbol{\delta \psi}_{\mathrm{MAP}} = \left( \mathbf{L}^{\mathrm{T}}_{\delta \psi} \mathbf{C}_{\mathrm{D}}^{-1} \mathbf{L}_{\delta \psi} + \mathbf{R}_{\delta \psi} \right)^{-1} \mathbf{L}^{\mathrm{T}}_{\delta \psi} \mathbf{C}_{\mathrm{D}}^{-1} \delta \boldsymbol{d}.
\label{eq:dpsi_only_mp_inv}
\end{equation}
Here, the regularisation matrix $\mathbf{R}_{\delta \psi}$ imposes a covariance prior on the pixelised potential correction $\boldsymbol{\delta \psi}$, thereby reducing the effective degrees of freedom and avoiding ill-posed solutions. The optimal regularisation $\mathbf{R}_{\delta \psi}$ can be determined by maximising the Bayesian evidence $\mathcal{E}$, whose formula was derived by \citet{Suyu2006_bayes} and reorganised by \citet{Dye2008}, as follows:
\begin{equation}
\begin{aligned}
\log\mathcal{E} = & -\frac{N_{\mathrm{d}}}{2}\log(2\pi) - \frac{1}{2}\log(\det \mathbf{C}_{\mathrm{D}}) \\
& + \frac{1}{2}\log(\det \mathbf{R}_{\delta \psi}) - \frac{1}{2} \delta \boldsymbol{\psi}^{\mathrm{T}}_{\mathrm{MAP}} \mathbf{R}_{\delta \psi} \delta \boldsymbol{\psi}_{\mathrm{MAP}} \\
& -\frac{1}{2} (\mathbf{L}_{\delta \psi} \delta \boldsymbol{\psi}_{\mathrm{MAP}} - \delta \boldsymbol{d})^{\mathrm{T}} \mathbf{C}_{\mathrm{D}}^{-1} (\mathbf{L}_{\delta \psi} \delta \boldsymbol{\psi}_{\mathrm{MAP}} - \delta \boldsymbol{d}) \\
& - \frac{1}{2} \log[\det(\mathbf{L}^{\mathrm{T}}_{\delta \psi} \mathbf{C}_{\mathrm{D}}^{-1} \mathbf{L}_{\delta \psi} + \mathbf{R}_{\delta \psi})].
\end{aligned}
\label{eq:dpsi_only_ev_eq}
\end{equation}
A more thorough description of the full Bayesian framework can be found in Appendix~\ref{sec:appdx_C}.

\subsection{Simultaneous Inversion of Source and Potential Perturbations}
\label{sec:inv_dpsi_src}
In Section~\ref{sec:inv_dpsi_only}, we fix the source light to the result given by the macro model, fitting only for $\boldsymbol{\delta \psi}$ using equation~\ref{eq:dpsi_linear_response}. However, this process does not account for the covariance between the source light and $\boldsymbol{\delta \psi}$. It has been shown that the source light model can absorb the image residual left by the imperfect lens mass model \citep[e.g.][]{Vernardos2022, Nightingale2024}. Therefore, a better approach is to fit the source light and $\boldsymbol{\delta \psi}$ simultaneously, allowing them to compete to reconstruct the observed lensing image. As demonstrated by \cite{Koopmans2005, Vegetti2009_method}, this can be achieved with simple manipulations on equation~\ref{eq:dpsi_linear_response}. Suppose we use a pixelized source model for the macro model, then the image residual left by the macro model, which appears on the left-hand side of equation~\ref{eq:dpsi_linear_response}, is equal to
\begin{equation}
\delta \boldsymbol{d} = \boldsymbol{d} - \boldsymbol{B} \mathbf{L_s}(\boldsymbol{\psi}_p) \boldsymbol{s}.
\end{equation}
Each element $d_j$ in the vector $\boldsymbol{d}$ represents the brightness value of the lensing image at each pixel. Each element $s_i$ in the vector $\boldsymbol{s}$ represents the brightness value at each source pixel. $\mathbf{L_s}(\boldsymbol{\psi}_p)$ is the lens mapping matrix with dimensions $[N_d, N_s]$, which, when acting on $\boldsymbol{s}$, generates the lensed source image. The value of $\mathbf{L_s}(\boldsymbol{\psi}_p)$ depends on the macro model's lens mass model (denoted by its lensing potential $\boldsymbol{\psi}_p$) and the interpolation scheme of the pixelized source model. For a more thorough illustration of how to construct this matrix $\mathbf{L_s}(\boldsymbol{\psi}_p)$, refer to Appendix B of \citet{Treu2004}. On the right-hand side of equation~\ref{eq:dpsi_linear_response}, the construction of the matrix $\mathbf{L_{\delta \psi}}$ requires knowledge of the source brightness distribution $S$, which can be obtained by interpolating the pixelized source reconstruction (denoted by $\boldsymbol{s}_p$), given by the macro model. Reorganizing equation~\ref{eq:dpsi_linear_response}, we have
\begin{equation}
\begin{aligned}
\boldsymbol{d} &= \boldsymbol{B} \mathbf{L_s}(\boldsymbol{\psi}_p) \boldsymbol{s} + \boldsymbol{B} \mathbf{L_{\delta \psi}}(\boldsymbol{s}_p) \boldsymbol{\delta \psi} + \boldsymbol{n} \\
&= \boldsymbol{B} \mathbf{L}(\boldsymbol{\psi}_p, \boldsymbol{s}_p) \boldsymbol{r} + \boldsymbol{n}.
\end{aligned}
\label{eq:dpsi_src_linear_response}
\end{equation}
Where we introduce the block matrices
\begin{equation}
\mathbf{L}(\boldsymbol{\psi}_p, \boldsymbol{s}_p) \equiv \left(\mathbf{L_s}(\boldsymbol{\psi}_p) \mid \mathbf{L_{\delta \psi}}(\boldsymbol{s}_p)\right),
\label{eq:L_mat}
\end{equation}
and 
\begin{equation}
\boldsymbol{r} \equiv \left(\begin{array}{c}
\boldsymbol{s} \\
\delta \boldsymbol{\psi}
\end{array}\right).
\label{eq:r_vec}
\end{equation}
It is immediately clear that the matrix equation~\ref{eq:dpsi_src_linear_response} is linear; therefore, all the mathematical frameworks shown in Section~\ref{sec:inv_dpsi_only} to solve Equation~\ref{eq:dpsi_linear_response} can be applied to Equation~\ref{eq:dpsi_src_linear_response}. All we need to do is to replace $\delta \boldsymbol{d}$ with $\boldsymbol{d}$, $\mathbf{L_{\delta \psi}}$ with $\mathbf{L}$, and $\boldsymbol{\delta \psi}$ with $\boldsymbol{r}$ \citep[see also][]{Vernardos2022}. One issue worth noting is that the regularisation prior $\{\boldsymbol{g}_{r}, \boldsymbol{\xi}_{r}\}$ for the vector $\boldsymbol{r}$ can be split into two independent parts: the source light regularisation $\{\boldsymbol{g}_{s}, \boldsymbol{\xi}_{s}\}$ and the lensing potential perturbation regularisation $\{\boldsymbol{g}_{\delta \psi}, \boldsymbol{\xi}_{\delta \psi}\}$. The regularisation functions for both $\boldsymbol{s}$ and $\delta \boldsymbol{\psi}$ also take a quadratic form, expressed as
\begin{equation}
\begin{aligned}
E_{s}(\boldsymbol{s}) &= \frac{1}{2} \boldsymbol{s}^{\mathrm{T}} \mathbf{R_{s}} \,\boldsymbol{s}, \\
E_{\delta \psi}(\delta \boldsymbol{\psi}) &= \frac{1}{2} \delta \boldsymbol{\psi}^{\mathrm{T}} \mathbf{R_{\delta \psi}} \delta \,\boldsymbol{\psi}.
\end{aligned}
\end{equation}
Here, $\mathbf{R_{s}}$ and $\mathbf{R_{\delta \psi}}$ represent the Hessians of the regularisation functions for the source ($E_{s}$) and the lensing potential perturbation ($E_{\delta \psi}$), respectively. The regularisation function of $\boldsymbol{r}$ is
\begin{equation}
\begin{aligned}
E_{r}(\boldsymbol{r}) &= E_{s}(\boldsymbol{s}) + E_{\delta \psi}(\delta \boldsymbol{\psi})\\
&= \frac{1}{2} \boldsymbol{r}^{\mathrm{T}} \mathbf{R_{r}} \boldsymbol{r}
\end{aligned}
\end{equation}
whose Hessian $\mathbf{R_{r}}$ can be further decomposed as
\begin{equation}
\mathbf{R}_{r} = \left(\begin{array}{cc}
\mathbf{R}_{s} & \mathbf{0} \\
\mathbf{0} & \mathbf{R}_{\delta \psi}
\end{array}\right).
\end{equation}
We note that solving the unknown perturbations in the lensing potential, whether based on equation~\ref{eq:dpsi_linear_response} or \ref{eq:dpsi_src_linear_response}, requires a non-zero source gradient (embedded in the $\boldsymbol{D_s}$ matrix). This implies that the potential correction method is, in principle, only applicable in the annular region where lensed emission is present. In practice, it is possible to use a larger mask that includes more regions without significant lensed emission and to let regularisation forcibly remove ill-posed solutions. However, as discussed in Section~\ref{sec:discuss_adpt_reg}, this approach tends to be sub-optimal.

\begin{table*}
    \renewcommand{\arraystretch}{1.2}
    \centering
    \begin{tabular}{|c|c|c|c|}
        \hline
        \multicolumn{4}{|c|}{\textbf{Free-Form Source Model}} \\ \hline
        \multirow{4}{*}{\textbf{Pixelization}} & \texttt{Overlay}                          & \texttt{Shape}                             & Fixed                  \\ \cline{2-4} 
                                                  & \multirow{3}{*}{\texttt{KMeans}}        & \texttt{Pixels}                           & Fixed                  \\ \cline{3-4} 
                                                  &                                          & \texttt{Weight Floor}                     & $\mathcal{L}(10^{-5}, 1.0)$    \\ \cline{3-4} 
                                                  &                                          & \texttt{Weight Power}                     & $\mathcal{U}(0.0, 5.0)$        \\ \hline
        \multirow{4}{*}{\textbf{regularisation}} & \texttt{ConstantSplit}                   & \texttt{Coefficient}                      & $\mathcal{L}(10^{-6}, 10^6)$   \\ \cline{2-4} 
                                                  & \multirow{3}{*}{\texttt{AdaptiveBrightnessSplit}} & \texttt{Inner Coefficient}              & $\mathcal{L}(10^{-6}, 10^6)$   \\ \cline{3-4} 
                                                  &                                          & \texttt{Outer Coefficient}                & $\mathcal{L}(10^{-6}, 10^6)$   \\ \cline{3-4} 
                                                  &                                          & \texttt{Signal Scale}                     & $\mathcal{U}(0.0, 1.0)$        \\ \hline
        \textbf{Interpolation}                   & \texttt{VoronoiNN}                       & \multicolumn{2}{c|}{\textbf{N/A}}         \\ \hline
        \multicolumn{4}{|c|}{\textbf{Free-Form Model for Lensing Potential Perturbations}} \\ \hline
        \textbf{Pixelization}                    & \texttt{RegularDpsiMesh}                 & \texttt{Factor}                           & 2                               \\ \hline
        \multirow{3}{*}{\textbf{regularisation}} & \multirow{3}{*}{\texttt{MaternKernel}}  & \texttt{Coefficient}                      & $\mathcal{L}(10^{-6}, 10^6)$   \\ \cline{3-4} 
                                                  &                                          & \texttt{Scale} [\arcsec]                            & $\mathcal{L}(10^{-4}, 10^3)$    \\ \cline{3-4} 
                                                  &                                          & \texttt{Nu}                               & $\mathcal{U}(0.5, 10.0)$        \\ \hline
        \textbf{Interpolation}                   & \text{Bilinear Interpolation}            & \multicolumn{2}{c|}{\textbf{N/A}}         \\ \hline
    \end{tabular}
    \caption{Settings for the Free-Form Models of the Source and Lensing Potential Perturbations. From left to right, the table presents the components of the free-form model (pixelization, regularisation, or interpolation), the specific code modules employed, the parameter names of the code modules, and their prior settings. $\mathcal{U}$ and $\mathcal{L}$ denote uniform and log-uniform priors, respectively.}
    \label{table:prior_setting}
\end{table*}

\subsection{Configurations for the Free-Form Model}
\label{sec:reg_func}
The free-form model discussed in this work represents model quantities using a pixelized field defined on a set of discretised grids, where each pixel's value is a free parameter. A free-form model includes three key components: pixelization, which determines the construction of the discretised grids; regularisation, which specifies how the field values across different pixels are correlated or smoothed; and the interpolation scheme, which describes how to map a pixelized field to a continuous one. In Sections~\ref{sec:reg_func_src} and \ref{sec:reg_func_dpsi}, we present the configurations of free-form models for the source light and lensing potential perturbations, respectively. Their specific settings are summarised in Table~\ref{table:prior_setting}.

\subsubsection{Configurations for Source Light}
\label{sec:reg_func_src}
In this work, we employ the \texttt{Overlay} model in \texttt{PyAutoLens} as the default pixelization scheme for the free-form source model. The \texttt{Overlay} scheme begins by overlaying a rectangular grid of size \(n_x \times n_y\) pixels within the modelling area defined by the mask\footnote{The mask identifies the region of the lensing image \textbf{included} in the lens model.}. Subsequently, the grid of unmasked pixels is mapped to the source plane according to the current lens mass model, forming the source grid. The values of \(n_x\) and \(n_y\) are free parameters for the \texttt{Overlay} model, and their optimal values can be determined using Bayesian evidence. When the structure of a background source in a lens system is compact and clumpy, the \texttt{KMeans} model in \texttt{PyAutoLens} can be used to create grids for the free-form source model. This ensures that more source pixelisation generating points cluster in regions where the source is brighter, achieving the resolution needed to fit the lensing image. Specifically, the \texttt{KMeans} model uses the emission from the lensed source as a weight (by applying the k-means clustering algorithm) to randomly select $N_{s}$ source pixels within the unmasked modelling area. This approach naturally generates a greater number of source pixels in regions with brighter lensed source emission. The grids on the image plane are then traced back to the source based on the current lens mass model to form the source grids. $N_{s}$ can be either a free model parameter or fixed; the default setting in this work is to set $N_{s}$ to half of the total number of image pixels within the unmasked annular region enclosing the lensed images.

The \texttt{ConstantSplit} model in \texttt{PyAutoLens} is employed as the default regularisation choice for the free-form source model. Essentially, this model regularises the brightness value of arbitrary source pixels, $s_j$, with all of their neighbouring pixels in a gradient fashion. Specifically, the regularisation function follows the form:
\begin{equation}
\begin{aligned}
E_{s}(\boldsymbol{s}) = \sum_{j=1}^J & \lambda_j^s \left\{\left[s_j - \tilde{s}\left(x_j + l_j, y_j + l_j\right)\right]^2\right. \\
& + \left[s_j - \tilde{s}\left(x_j - l_j, y_j + l_j\right)\right]^2 \\
& + \left[s_j - \tilde{s}\left(x_j + l_j, y_j - l_j\right)\right]^2 \\
& + \left[s_j - \tilde{s}\left(x_j - l_j, y_j - l_j\right)\right]^2\}.
\end{aligned}
\end{equation}
Here, $\tilde{s}$ represents a continuous expression of the pixelized source model, derived using natural neighbour interpolation \citep{Sibson1981} under the Voronoi tessellation for the source grid. $(x_j, y_j)$ denote the grid coordinates of the $j$-th source pixel, while $l_j$ is the ``cross size'' of the $j$-th source pixel, calculated as the square root of the area of the $j$-th source pixel's Voronoi cell. The factor $\lambda_j^s$ determines the overall regularisation (or smoothing) strength imposed on the $j$-th source pixel; for the \texttt{ConstantSplit} model, $\lambda_j^s$ is a universal value across all source pixels. In cases where the source shows a high spatial variance in brightness distribution, regularisation schemes that allow for more dynamic range are needed. For faint regions on the source plane, we impose a higher regularisation strength to suppress noise, while for bright regions, we impose a lower regularisation to give the free-form model more freedom to reconstruct detailed morphological features of the source. The \texttt{AdaptiveBrightnessSplit} model in \texttt{PyAutoLens} provides such functionality. Specifically, this is done by directly ray-tracing the lensed source emission back to the source plane based on the current lens mass model and calculating the average brightness value $A_j$ for the $j$-th source pixel. The $A_j$ value is then mapped to the regularisation strength factor $\lambda_j^s$, such that a lower $A_j$ value corresponds to a higher $\lambda_j^s$ value.

For more detailed information on various source pixelization, regularisation, and interpolation schemes provided by \texttt{PyAutoLens}, we refer readers to \citet{PyAutoLens_2, PyAutoLens_1, Nightingale2024, He2024_mge}.

\subsubsection{Configurations for Lensing Potential Perturbations}
\label{sec:reg_func_dpsi}
Solving the unknown lensing potential perturbation with Equation~\ref{eq:dpsi_linear_response} or \ref{eq:dpsi_src_linear_response} requires knowledge of non-vanishing source gradient information, meaning the potential correction method only works in an annular region that tightly encloses the lensed source images. The pixelization grid for the free-form model of lensing potential perturbation is thus defined within this annular region, using a regular grid with a pixel size $N_{\delta \psi}$ times larger than that of the native lensing image\footnote{This functionality is supported by the \texttt{RegularDpsiMesh} module in our code, where the \texttt{factor} parameter corresponds to $N_{\delta \psi}$.}, where \(N_{\delta \psi} = 1, 2, \ldots\). The default $N_{\delta \psi}$ value used in this work is 2, which strikes a balance between computational efficiency and the accuracy required to recover the small mass perturbations induced by subhalos.

The Mat\'ern regularisation \citep{stein1999_matern, Rasmussen2006_gpr, matern2013spatial} is employed to control the smoothness of the pixelized lensing potential perturbation. For any two pixels of the lensing potential perturbation, denoted by their coordinates \((x_m, y_m)\) and \((x_n, y_n)\), the correlation between them can be described by
\begin{equation}
C_{ij}(\rho, \nu) = \frac{2^{1-\nu}}{\Gamma(\nu)}\left(\sqrt{2 \nu} \frac{d_{ij}}{\rho}\right)^\nu K_\nu\left(\sqrt{2 \nu} \frac{d_{ij}}{\rho}\right),
\label{eq:matern_kernel}
\end{equation}
where $\Gamma$ is the gamma function, and $K_\nu$ is the modified Bessel function of the second kind. \(d_{ij} = \sqrt{(x_m-x_n)^2 + (y_m-y_n)^2}\) represents the distance between these two pixels. The parameter \(\rho\) defines the characteristic smoothing scale, while \(\nu\) governs the order of smoothing, ensuring that the pixelized model reconstruction is \(\left\lceil \nu \right\rceil - 1\) times differentiable. The Exponential and Gaussian kernels used in \citet{Vernardos2022} are special cases of the Mat\'ern kernel, with $\nu = 0.5$ and $\nu = +\infty$, respectively. The Hessian (\(\mathbf{R}_{\delta \psi}\)) of the regularisation function, using the Mat\'ern scheme, is defined as follows:
\begin{equation}
\mathbf{R}_{\delta \psi}(\lambda^{\delta \psi}, \rho, \nu) = \lambda^{\delta \psi} \mathbf{C}_{\delta \psi}^{-1}(\rho, \nu),
\end{equation}
where \(\mathbf{C_{\delta \psi}}\) is the covariance matrix with elements \(C_{ij}\), and \(\lambda^{\delta \psi}\) controls the overall strength of the regularisation. Bilinear interpolation is employed to interpret the pixelized lensing potential perturbations as a continuous field.
\section{Overview of Strong Lensing Data}
\label{sec:dataset}
We present an overview of the galaxy-galaxy strong lensing data used in this work. Section~\ref{sec:data_mock} outlines the image simulation settings for the mock lenses employed to demonstrate the capability of our method. In Section~\ref{sec:data_0946}, we describe the HST observational data for the lens system SLACS0946+1006, which is re-analysed using our new potential correction code.

\begin{table*}
    \renewcommand{\arraystretch}{1.2}
    \centering
    \begin{tabular}{|c|c|c|}
        \hline
        \multicolumn{3}{|c|}{\textbf{Main Lens's Mass Distribution}} \\ \hline
        \multirow{6}{*}{\textbf{Elliptical Power Law}} & Centre (y, x) $[\arcsec]$                   & (0.0, 0.0) \\ \cline{2-3}
                                                       & Axis Ratio ($q$)                            & 0.9       \\ \cline{2-3}
                                                       & Position Angle ($\phi$) $[\degr]$           & 0.0       \\ \cline{2-3}
                                                       & Einstein Radius ($\theta_\mathrm{E}$) $[\arcsec]$ & 1.8   \\ \cline{2-3}
                                                       & Density Slope ($\gamma$)                    & 2.0       \\ \cline{2-3}
                                                       & Redshift                                     & 0.2       \\ \hline
        \multicolumn{3}{|c|}{\textbf{Source Brightness Distribution}} \\ \hline
        \multirow{11}{*}{\textbf{Two Elliptical Gaussian Models}} & \multirow{2}{*}{Centre (y, x) $[\arcsec]$} & (0.0, 0.0) \\ \cline{3-3}
                                                       &                                             & (0.0, 0.4) \\ \cline{2-3}
                                                       & \multirow{2}{*}{Axis Ratio}                & 0.6       \\ \cline{3-3}
                                                       &                                             & 0.4       \\ \cline{2-3}
                                                       & \multirow{2}{*}{Position Angle $[\degr]$}   & 45.0      \\ \cline{3-3}
                                                       &                                             & 135.0     \\ \cline{2-3}
                                                       & \multirow{2}{*}{Standard Deviation ($\sigma$) $[\arcsec]$} & 0.15 \\ \cline{3-3}
                                                       &                                             & 0.1       \\ \cline{2-3}
                                                       & \multirow{2}{*}{Intensity $[e^{-1}s^{-1}]$} & 5.0       \\ \cline{3-3}
                                                       &                                             & 3.0       \\ \cline{2-3}
                                                       & Redshift                                     & 0.6       \\ \hline
        \multicolumn{3}{|c|}{\textbf{Lensing Potential Perturbations}} \\ \hline
        \multirow{4}{*}{\textbf{Spherical NFW Subhalo}} & Centre (y, x) $[\arcsec]$                   & (1.81, 0.0) \\ \cline{2-3}
                                                       & Mass ($M_{200}$) $[M_{\odot}]$              & $10^{10}$  \\ \cline{2-3}
                                                       & Concentration ($c$)                         & 11.60      \\ \cline{2-3}
                                                       & Redshift                                     & 0.2        \\ \hline
        \multirow{2}{*}{\textbf{Shear}}                   & Magnitude ($\gamma^\mathrm{ext}$)           & 0.01       \\ \cline{2-3}
                                                       & Orientation ($\phi^\mathrm{ext}$) $[\degr]$  & 45.0       \\ \hline
        \multirow{3}{*}{\textbf{Multipole}}               & Order ($m$)                                 & 4          \\ \cline{2-3}
                                                       & Normalization ($a_m$)                       & 0.0569     \\ \cline{2-3}
                                                       & Orientation ($\phi_m$) $[\degr]$             & 0.0        \\ \hline
        \multirow{2}{*}{\textbf{Gaussian Random Field}}   & Normalization ($\log[A_\mathrm{GRF}]$)     & $-6.0$ $\left(-4.35^{0.03}_{-0.03}\right)$ \\ \cline{2-3}
                                                       & Power Spectrum Slope ($\beta_\mathrm{GRF}$) & 6.0 $\left(2.72^{0.05}_{-0.05}\right)$        \\ \hline
    \end{tabular}
    \caption{Simulation setup for generating mock lensing images. The main lens mass and source light properties were chosen to match typical values from the SLACS sample. Mass perturbations were set to astrophysically plausible values. For the Gaussian random field, the values on the left correspond to the ground-truth power spectrum properties used to generate the lensing potential perturbation across the field of view. In contrast, the values in parentheses represent the power spectrum properties computed within the annular unmasked region where the lens modelling is performed. These properties were derived by fitting a power-law model to the one-dimensional power spectrum of lens potential perturbations calculated within this region, with superscripts and subscripts indicating the $68\%$ confidence interval.}
    \label{table:mock_setting}
\end{table*}

\subsection{Mock Lensing Data}
\label{sec:data_mock}

\begin{figure*}
	\includegraphics[width=\textwidth]{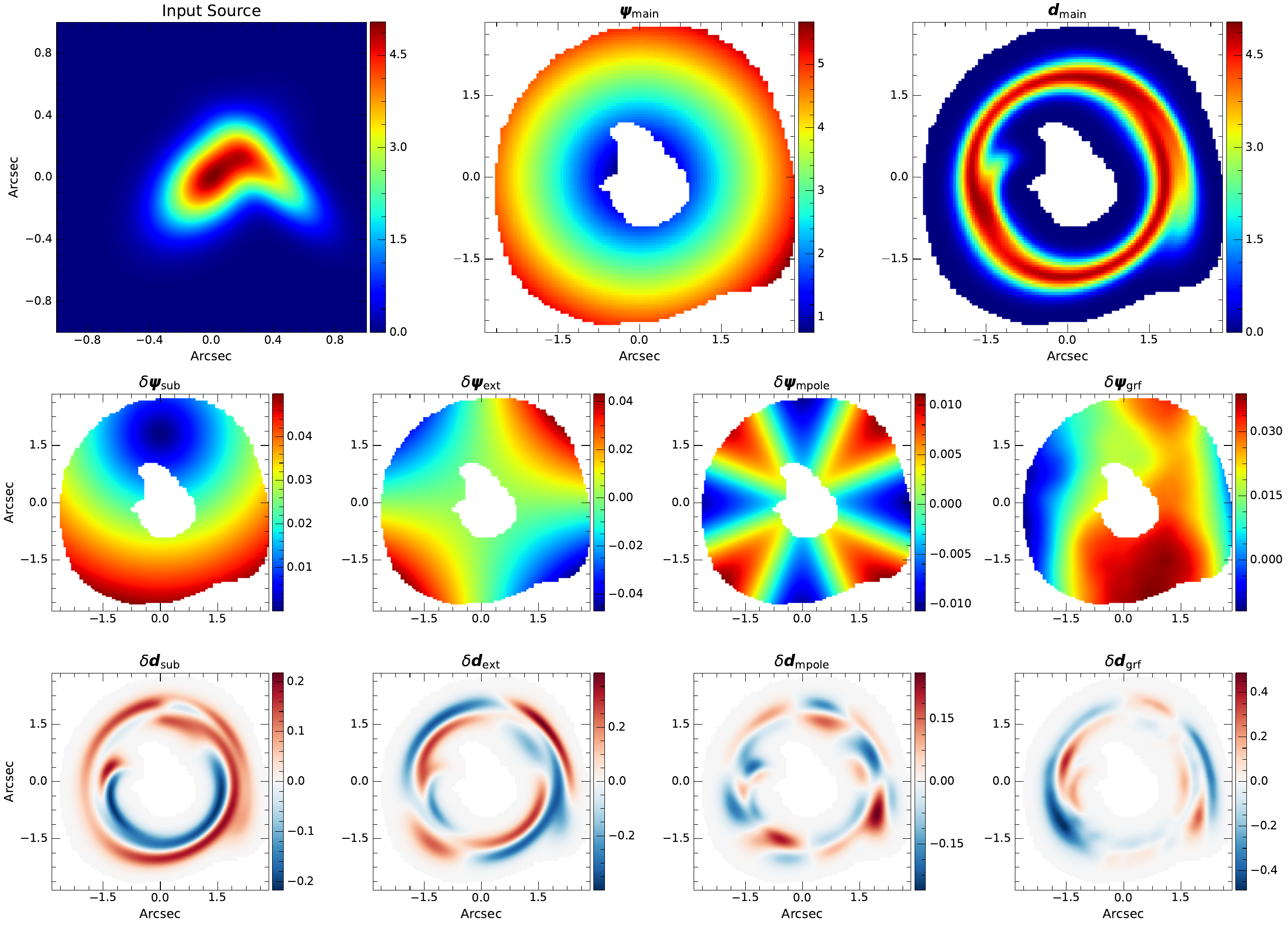}
        \caption{Mock datasets used in this work (see Table~\ref{table:mock_setting} for specific parameter values). The first row depicts the intrinsic source image, the lensing potential of the main lens, and the resulting lensed image. The second row shows the lensing potential maps of various perturbative signals: a $10^{10}$~M$_{\odot}$ NFW subhalo, an external shear field, a $m=4$ multipole component, and a Gaussian random field. The third row presents the corresponding image perturbations induced by these lensing potential perturbations. Unless otherwise noted, lensing potentials are shown in units of $\text{arcsec}^2$, and lensing images are in units of $\mathrm{e}^- \ \mathrm{s}^{-1} \ \mathrm{pixel}^{-1}$.}
    \label{fig:mock_dataset}
\end{figure*}

We generate a set of mock lensing images that incorporate various perturbation signals in the lensing potential to evaluate the effectiveness of the potential correction code developed in this study. To control for variables, we ensure that the mass of the main lens galaxy and the source light remain consistent across all mock datasets, allowing only the perturbation signals in the lensing potential to vary. The mass distribution of the main lens galaxy is described by the elliptical power-law model \citep{EPL_model}, whose dimensionless surface mass density (i.e.\ convergence) profile is expressed as
\begin{equation}
\kappa(r) = \frac{3 - \gamma}{2} \left( \frac{\theta_\mathrm{E}}{r} \right)^{\gamma - 1}.
\end{equation}
Here, \(1 < \gamma < 3\) represents the slope of the deprojected volume density, \(\theta_\mathrm{E}\) denotes the Einstein radius, and \(r\) signifies the elliptical radial distance from the centre of the lens galaxy. The elliptical radial distance is defined as \(r(x,y) = \sqrt{x^2 q + \frac{y^2}{q}}\), where \(q\) is the minor-to-major axis ratio, and \(\phi\) is the position angle measured counterclockwise from the positive x-axis. The parameters \(q\) and \(\phi\) are reparametrized into two ellipticity components, \(\epsilon_1\) and \(\epsilon_2\), defined as
\begin{equation}
\epsilon_1 = \frac{1 - q}{1 + q} \sin 2\phi, \quad \epsilon_2 = \frac{1 - q}{1 + q} \cos 2\phi,
\end{equation}
to facilitate more efficient model fitting \citep{Birrer2015}. The surface brightness distribution of the source is characterised by the superposition of two elliptical Gaussian models, reflecting the irregular morphology observed in real lenses \citep[e.g.][]{Bolton2008, Etherington2022}.

Four popular types of lensing potential perturbation signals, frequently used to evaluate the performance of free-form models \citep[e.g.][]{Vernardos2022, Galan2022_wavelet, Biggio2023}, are employed to construct mock datasets and test our potential correction code. These include: a spherical Navarro-Frenk-White (NFW) profile representing subhalos within the lens galaxy \citep{NFW_model}, an external shear field caused by neighbouring galaxies or the cosmic large-scale structure \citep{Keeton1997}, a multipole perturbation induced by galaxy interactions such as mergers \citep{Hao2006}, and a Gaussian Random Field (GRF), which is expected to capture the collective lensing effects of numerous line-of-sight halos \citep{Chatterjee2018, Bayer2023}. The NFW model has a volume mass density profile of
\begin{equation}
\rho = \frac{\rho_\mathrm{s}}{(r / r_\mathrm{s})(1 + r / r_\mathrm{s})^2},
\end{equation}
where $\rho_\mathrm{s}$ is the normalization factor and $r_\mathrm{s}$ is the scale radius. $\rho_\mathrm{s}$ and $r_\mathrm{s}$ are typically reparametrized into the halo mass $M_\mathrm{200}$ and concentration $c = r_\mathrm{200} / r_\mathrm{s}$, where $M_\mathrm{200}$ is the mass enclosed within the radius $r_\mathrm{200}$, within which the average density is 200 times the critical density of the Universe. Here, we assume that NFW halos follow the mass-concentration relation of \citet{Ludlow2016}. Thus, the NFW model is solely quantified by $M_\mathrm{200}$. The external shear is parameterized as two elliptical components ($\gamma_1^\mathrm{ext}$, $\gamma_2^\mathrm{ext}$), whose shear magnitude $\gamma^\mathrm{ext}$ and orientation $\phi^\mathrm{ext}$ measured counter-clockwise from the north are given by
\begin{equation}
\gamma^\mathrm{ext} = \sqrt{(\gamma_1^\mathrm{ext})^2 + (\gamma_2^\mathrm{ext})^2}, \quad \tan 2\phi^\mathrm{ext} = \frac{\gamma_2^\mathrm{ext}}{\gamma_1^\mathrm{ext}}.
\end{equation}
The lensing potential of the shear model is usually expressed in polar coordinates $(\theta, \varphi)$ by
\begin{equation}
\psi^\mathrm{ext}(\theta, \varphi) = -\frac{1}{2} \gamma^\mathrm{ext} \theta^2 \cos 2(\varphi - \phi^\mathrm{ext}).
\end{equation}
The lensing potential of the multipole model \citep{Xu2015,Vyvere2022,Lange2024} is expressed as
\begin{equation}
\psi^\text{mpole}(\theta) = \frac{r}{1 - m^2} a_m \cos(m \phi - m \phi_m),
\end{equation}
where \(a_m\) represents the normalization factor, and \(m\) and \(\phi_m\) denote the multipole order and orientation, respectively \citep[c.f.][]{Congdon2005,Taylor2017,Etherington2024,Paugnat2025}. The characteristics of a GRF are defined by its isotropic power spectrum, which follows a power-law relation:
\begin{equation}
P(k) = A_\mathrm{GRF} k^{-\beta_\mathrm{GRF}},
\end{equation}
where $A_\mathrm{GRF}$ is the normalization factor, $k$ is the wave number, and $\beta_\mathrm{GRF}$ is the slope. By specifying particular values for $A_\mathrm{GRF}$ and $\beta_\mathrm{GRF}$, a random realisation of a GRF-like lensing potential field can be generated using
\begin{equation}
\psi^\mathrm{GRF} = \mathcal{F}^{-1}(P(k), \epsilon),
\label{eq:grf_ivf}
\end{equation}
where $\epsilon$ represents the random number seed, and $\mathcal{F}^{-1}$ denotes the inverse Fourier transform. We employ the \texttt{powerbox} software \citep{Murray2018} to perform the operations described in Equation~\ref{eq:grf_ivf}.

The specific parameter settings used to generate the mock datasets are summarised in Table~\ref{table:mock_setting}, with the corresponding lensing images presented in Figure~\ref{fig:mock_dataset}. Since the potential correction method only works on an annular region encircling the lensed images, annular masks are automatically generated by identifying connected regions in the image where the pixel’s signal-to-noise ratio (SNR) is greater than 3, then dilating these regions by 10 pixels. Our mock datasets exhibit image quality comparable to that of the SLACS lens samples \citep{Bolton2008}, with a pixel size of $0.05\arcsec$ and a Gaussian PSF characterised by a $\sigma$ value of $0.05\arcsec$. The sky background level is set to $0.1\,\mathrm{e^{-}\,s^{-1}}$, and the lensing images are subjected to exposures of 840 seconds. The maximum SNR (in the brightest pixel) of the lensed arc in our mock datasets is approximately 50, roughly corresponding to the brightest systems in the SLACS sample. For simplicity, all our mock lensing images omit the lens galaxy's light during the simulation. This simplification does not compromise the generality of this work, as the lens galaxy's light can be pre-subtracted before applying the potential correction method.

\subsection{Lensing Data for SLACS0946+1006}
\label{sec:data_0946}

\begin{figure}
	\includegraphics[width=\columnwidth]{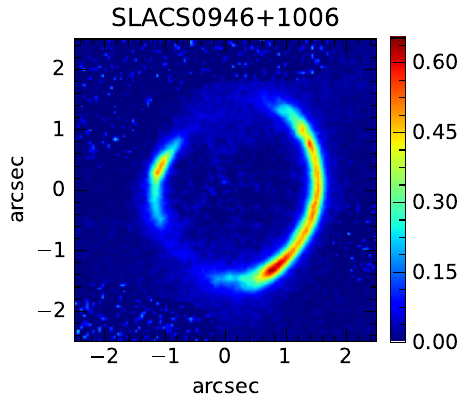}
        \caption{Lens-light subtracted image of SLACS0946+1006 in the HST F814W band, using the multiple Gaussian expansion model presented in \citet{He2024_mge}.}
    \label{fig:0946_dataset}
\end{figure}

SLACS0946+1006 is a compound lens system with a rare triple source-plane configuration, characterized by a lens redshift of $z_l = 0.222$ and source redshifts of $z_{s1} = 0.609$, $z_{s2} = 2.035$, and $z_{s3} = 5.975$ \citep{Smith2021}. This system was initially pre-selected as a lens candidate from early-type galaxies in the SDSS survey, which exhibited abnormal emission lines potentially originating from high-redshift background sources \citep{Bolton2006}. The ACS camera on board the HST was subsequently employed to confirm the strong lensing nature of this system. The SLACS0946+1006 data analysed in this work were obtained using the ACS F814W band and then drizzled to \(0.05 \arcsec \text{pixel}^{-1}\), following the reduction procedures described in \citet{Bolton2008}. The PSF was determined using the \texttt{TinyTim} software. For simplicity, we masked the outer Einstein ring, using only the innermost Einstein ring for potential correction. The lens light was pre-subtracted using the model results from \citet{He2024_mge}, which employs the Multipole Gaussian Expansion (MGE) model to represent the lens light; combined with a pixelized source model, the lens and source light are fit simultaneously to account for the covariance between the two. Subtracting the lens light using MGE typically outperforms multiple Sersic \citep{Sersic} fits in terms of image residuals. Figure~\ref{fig:0946_dataset} shows the lens-light-subtracted image of SLACS0946+1006.

\section{Result}
\label{sec:result}
In Section~\ref{sec:res_inv_dpsi_only}, we assume perfect knowledge of both the source light and the image residuals induced by input mass perturbations to solve for the pixelised lensing potential perturbations from these image residuals, using the framework outlined in Section~\ref{sec:inv_dpsi_only}. Although this test disregards the potential degeneracy between the source light and the lens mass, which could affect the performance of a free-form lens model, it provides valuable insights into the viability of our free-form approach, particularly the effectiveness of regularisation schemes in recovering various forms of lensing potential perturbations. In Section~\ref{sec:res_inv_dpsi_src}, we evaluate the performance of our potential correction algorithm in a more realistic context by simultaneously modelling the pixelised lensing potential perturbations and source light through direct fitting of the lensing images (using the framework presented in Section~\ref{sec:inv_dpsi_src}), thereby accounting for their covariance. Since the macro model typically absorbs some of the image signals caused by perturbations, it introduces biases in the estimated main lens mass and source light. To assess the impact of these biases on the performance of the potential correction algorithm, we conduct two tests: one initialises the potential correction with the true input models of the main lens mass and source light (Section~\ref{sec:res_inv_dpsi_src_true}), while the other uses the main lens mass and source light estimates derived from the macro model as the starting point (Section~\ref{sec:res_inv_dpsi_src_macro}). Finally, in Section~\ref{sec:res_0946}, we apply our method to the real HST lens system SLACS0946+1006 and compare our findings with previous studies.

\subsection{Inverting Only the Potential Perturbation}
\label{sec:res_inv_dpsi_only}

\begin{figure*}
	\includegraphics[width=\textwidth]{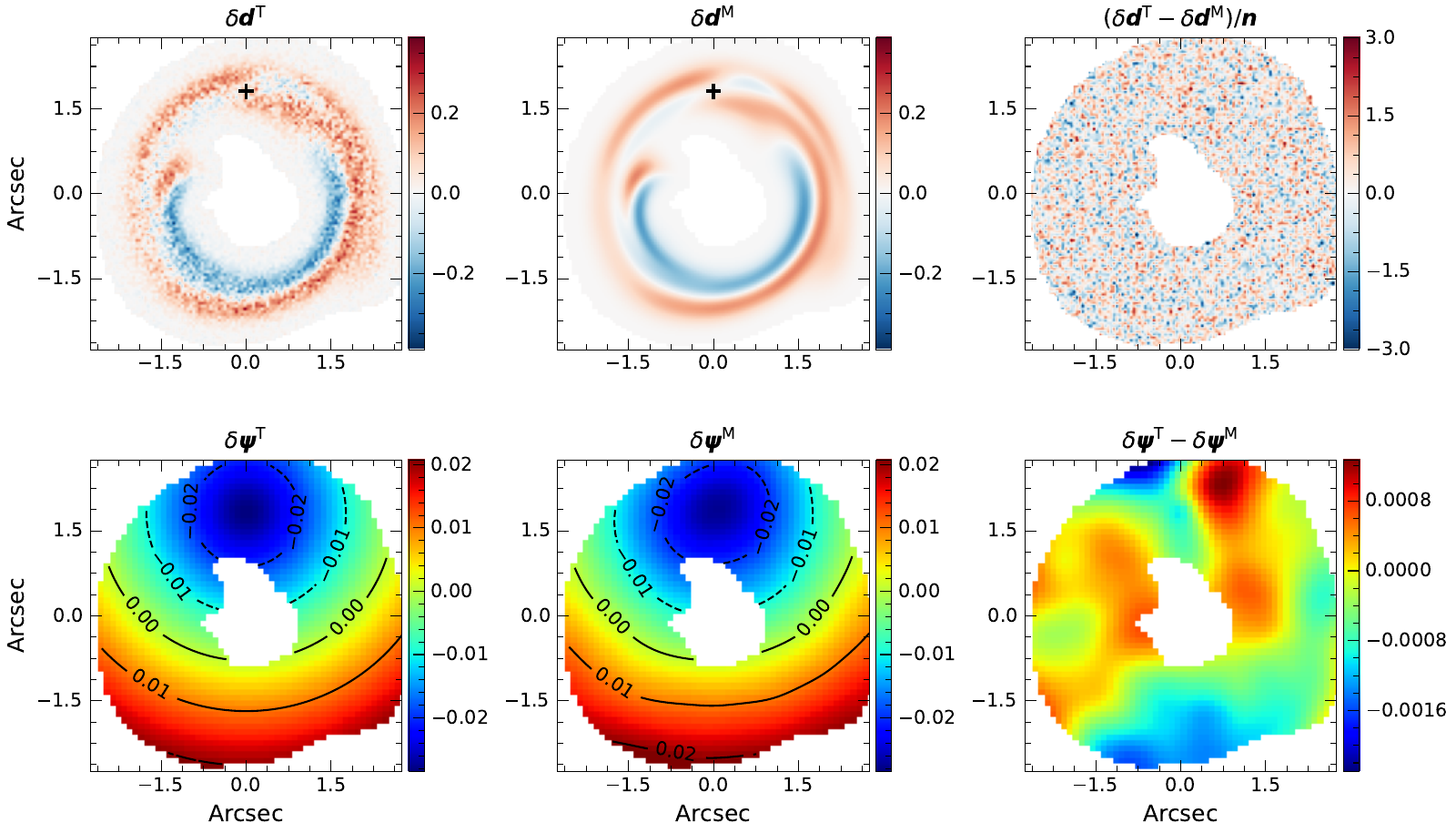}
        \caption{Reconstruction of the lensing potential perturbation induced by a $10^{10} M_{\odot}$ NFW subhalo using the Mat\'ern regularisation scheme, assuming that the true source and the image residuals induced by the perturber are perfectly known. \textit{Top-left}: Image residual induced by the $10^{10} M_{\odot}$ NFW subhalo. \textit{Top-middle}: Image residual reconstructed using the potential correction model. \textit{Top-right}: Difference between the top-left and top-middle panels, normalised by the noise map. \textit{Bottom-left}: Lensing potential map of the input perturber. \textit{Bottom-middle}: Lensing potential map of the model perturber, reconstructed using the potential correction model. \textit{Bottom-right}: Difference between the bottom-left and bottom-middle panels. The `plus' symbol in the top-left and top-middle panels marks the position of the input subhalo.}
    \label{fig:sub_dpsi_only_inv}
\end{figure*}

\begin{figure*}
	\includegraphics[width=0.6\textwidth]{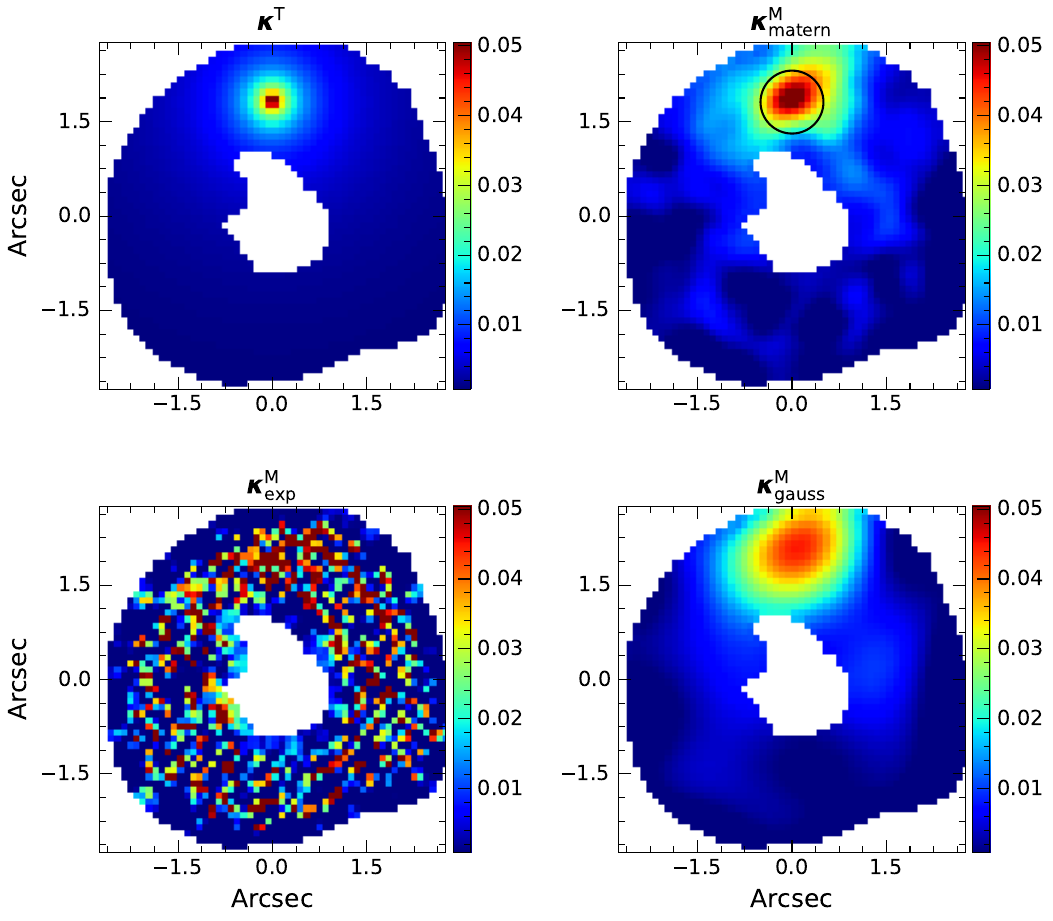}
        \caption{Reconstruction of the convergence perturbation induced by a $10^{10}\,M_{\odot}$ NFW subhalo using different regularisation schemes for the perturbative lensing potential. \textbf{Top-left}: Input convergence map of the $10^{10}\,M_{\odot}$ NFW subhalo. \textbf{Top-right}: Reconstructed convergence map using the potential correction method with Mat\'ern regularisation. The black circle (radius $0.5\arcsec$) indicates the region where a parametric NFW model is fit to the reconstructed lensing potential to assess consistency with the input perturbation. \textbf{Bottom-left}: Same as top-right, but with Exponential regularisation. \textbf{Bottom-right}: Same as top-right, but with Gaussian regularisation. Exponential and Gaussian regularisations tend to produce under- or over-smoothed solutions, whereas the Mat\'ern kernel yields a good reconstruction.}
    \label{fig:sub_dpsi_only_kappa}
\end{figure*}

\begin{figure*}
	\includegraphics[width=\textwidth]{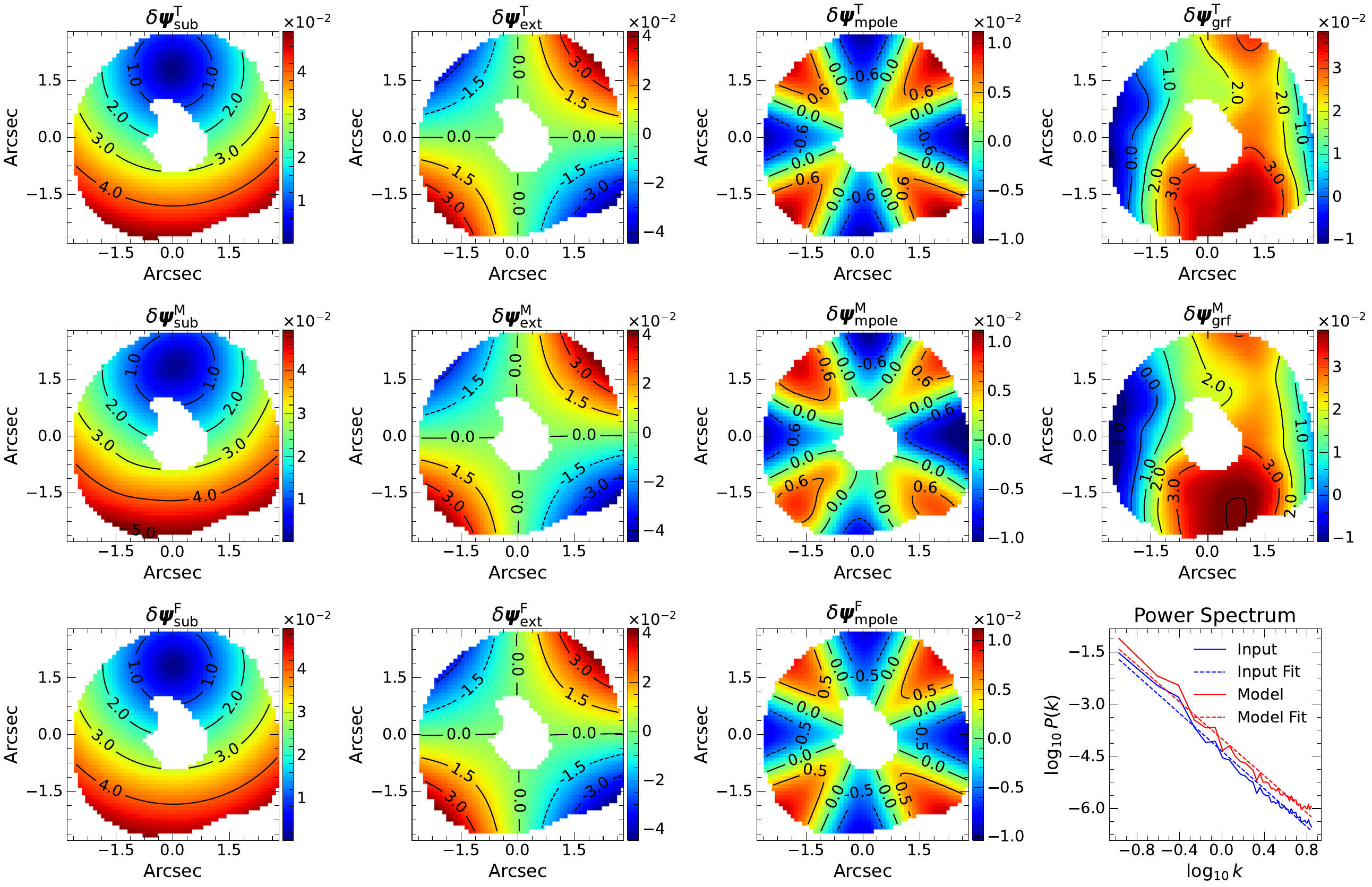}
        \caption{Comparison of reconstructed lensing potential perturbations induced by an NFW subhalo (left column), external shear (middle-left column), \(m_4\) multipole (middle-right column), and a Gaussian random field (GRF) (right column) with the ground truth. The first row presents the input lensing potential maps of perturbers. The second row shows the perturbative lensing potential maps derived from the potential correction model. The third row presents a quantitative analysis: for the NFW subhalo, external shear, and $m_4$ multipole, parametric models are fitted to the lensing potential map in the second row, with point estimates reported in Table~\ref{tab:diagnose_inversion}; for the GRF, the power spectra of the input (blue lines) and reconstructed (red lines) lensing potential maps are compared. Solid lines indicate power spectra calculated within the unmasked modelling region, while dashed lines represent power-law model fits.}
    \label{fig:dpsi_inv_diagnosis}
\end{figure*}

\begin{table*}
    \renewcommand{\arraystretch}{1.5}
    \centering
    \begin{tabular}{|c|c|c|c|c|}
        \hline
        \multicolumn{5}{|c|}{\textbf{Spherical NFW Subhalo}} \\ \hline
        \textbf{Parameter Names} & \textbf{$\delta \boldsymbol{\psi}$-Only} (\S~\ref{sec:inv_dpsi_only}) & \textbf{$(\delta \boldsymbol{\psi}, \boldsymbol{s})^\mathrm{Truth}$} (\S~\ref{sec:res_inv_dpsi_src_true}) & \textbf{$(\delta \boldsymbol{\psi}, \boldsymbol{s})^\mathrm{Macro}$} (\S~\ref{sec:res_inv_dpsi_src_macro}) & \textbf{Ground Truths} \\ \hline
        Centre (y, x) $[\arcsec]$ & $(1.8190^{0.1774}_{-0.1798}, 0.0154^{0.1703}_{-0.1797})$ & $(1.7714^{0.1716}_{-0.1668}, 0.0422^{0.1656}_{-0.1699})$ & $(1.7634^{0.0480}_{-0.0498}, 0.0455^{0.0473}_{-0.0474})$ & (1.81, 0.0) \\ \hline
        Mass ($\log_{10}[M_{200}/\mathrm{M}_\odot]$) & $9.9665^{0.7155}_{-0.6685}$ & $9.9151^{0.7237}_{-0.6488}$ & $9.5192^{0.8085}_{-0.3837}$ & 10.0 \\ \hline
        \multicolumn{5}{|c|}{\textbf{Shear}} \\ \hline
        Magnitude ($\gamma^\mathrm{ext}$) & $0.0101^{0.0008}_{-0.0008}$ & $0.0100^{0.0008}_{-0.0008}$ & $0.0110^{0.0001}_{-0.0001}$ & 0.01 \\ \hline
        Orientation ($\phi^\mathrm{ext}$) $[\degr]$ & $45.6705^{2.4147}_{-2.3474}$ & $44.5560^{2.4775}_{-2.3296}$ & $47.3874^{0.3598}_{-0.3756}$ & 45 \\ \hline
        \multicolumn{5}{|c|}{\textbf{Multipole}} \\ \hline
        Normalization ($a_m$) & $0.0507^{0.0006}_{-0.0007}$ & $0.0527^{0.0012}_{-0.0011}$ & $0.0478^{0.0010}_{-0.0011}$ & 0.0569 \\ \hline
        Orientation ($\phi_m$) $[\degr]$ & $-0.0208^{3.1411}_{-3.1396}$ & $-0.0081^{4.7032}_{-3.1458}$ & $-0.0251^{3.1449}_{-3.1467}$ & 0.0 \\ \hline
        \multicolumn{5}{|c|}{\textbf{Gaussian Random Field}} \\ \hline
        Normalization ($\log[A_\mathrm{GRF}]$) & $-4.00^{0.04}_{0.04}$ & $-4.48^{0.03}_{-0.03}$ & $-4.17^{0.04}_{-0.04}$ & $-4.35^{0.03}_{-0.03}$ \\ \hline
        Power Spectrum Slope ($\beta_\mathrm{GRF}$) & $2.66^{0.06}_{-0.06}$ & $2.70^{0.05}_{-0.05}$ & $2.56^{0.07}_{0.07}$ & $2.72^{0.05}_{-0.05}$ \\ \hline
    \end{tabular}
    \caption{Quantitative comparison of reconstructed and input lensing potential perturbations. For subhalo, shear, and multipole models, parametric models are fitted to the reconstructed perturbations to assess agreement with the input. For the Gaussian random field, the power spectrum of the perturbations is compared. The labels $\delta \boldsymbol{\psi}$-Only, $(\delta \boldsymbol{\psi}, \boldsymbol{s})^\mathrm{Truth}$, and $(\delta \boldsymbol{\psi}, \boldsymbol{s})^\mathrm{Macro}$ denote inversions of only the lensing potential perturbations (\S~\ref{sec:inv_dpsi_only}), inversions of both source and lensing potential perturbations starting from the exact true main lens mass and source light distributions (\S~\ref{sec:res_inv_dpsi_src_true}), and inversions of both source and lensing potential perturbations starting from the macro model fit (\S~\ref{sec:res_inv_dpsi_src_macro}), respectively. Superscripts and subscripts represent the 68\% credible interval.}
    \label{tab:diagnose_inversion}
\end{table*}

We evaluate the performance of our proposed correction method using the mock datasets described in Section~\ref{sec:data_mock}. We assume that the brightness distribution of the source and the image residuals induced by lensing potential perturbations (as illustrated in Figure~\ref{fig:mock_dataset}) are perfectly known a priori. This information, along with the input mass distribution of the main lens, is then utilised to construct the vector $\delta \boldsymbol{d}$ and the matrix $\boldsymbol{D}_{\mathrm{s}}$ in Equation~\ref{eq:dpsi_linear_response}. The framework presented in Section~\ref{sec:inv_dpsi_only} is employed to solve for the lensing potential perturbations $\boldsymbol{\delta \psi}$. Specifically, we apply the Mat\'ern regularisation discussed in Section~\ref{sec:reg_func_dpsi} to enforce smoothness constraints on $\boldsymbol{\delta \psi}$. Given an arbitrary set of hyperparameters denoted by $\boldsymbol{\xi}_{\delta \psi}$, the most probable solution $\boldsymbol{\delta \psi}_{\mathrm{mp}}$ can be derived using Equation~\ref{eq:dpsi_only_mp_inv}. Assuming that the prior $P(\boldsymbol{\xi}_{\delta \psi})$ for $\boldsymbol{\delta \psi}$ is noninformative (see Table~\ref{table:prior_setting}), the posterior distribution for the hyperparameters $P(\boldsymbol{\xi}_{\delta \psi} \mid \delta \boldsymbol{d}, \mathbf{L}_{\delta \psi}, \boldsymbol{g}_{\delta \psi})$ is determined by sampling the likelihood $P(\delta \boldsymbol{d} \mid \mathbf{L}_{\delta \psi}, \boldsymbol{\xi}_{\delta \psi}, \boldsymbol{g}_{\delta \psi})$, the specific form of which is given by Equation~\ref{eq:dpsi_only_ev_eq}. The optimal hyperparameter values determined via Bayesian evidence are summarised in Table~\ref{tab:optimal_hyper_params}.

The model result for mock data with an input NFW subhalo perturber is shown in Figure~\ref{fig:sub_dpsi_only_inv}. Our potential correction method successfully finds a solution for the lensing potential perturbations $(\delta\psi^{M})$ that produces a model image residual $(\delta d^{M})$ closely matching the observed residual $(\delta d^{T})$. Furthermore, $\delta\psi^{M}$ is highly consistent with the lensing potential map of the input perturber $(\delta\psi^{T})$.

To better visualise the structure of a localised subhalo, the convergence map is more informative than the lensing potential map. The top-left and top-right panels of Figure~\ref{fig:sub_dpsi_only_kappa} compare the convergence maps of the input perturber and the one derived by the potential correction model, respectively. Our potential correction method with Mat\'ern regularisation successfully reconstructs the mass distribution of the input NFW subhalo qualitatively. In contrast, using Exponential or Gaussian regularisation, as proposed in \citet{Vernardos2022}, leads to under- or over-smoothed solutions that fail to recover the input NFW subhalo, as shown in the bottom-left and bottom-right panels of Figure~\ref{fig:sub_dpsi_only_kappa}. This demonstrates the superiority of Matern regularisation in reconstructing localised subhalo perturbations.

To enable a more quantitative assessment of the potential correction method, we fit a parametric model to the perturbative lensing potential map derived by the potential correction method for mock lenses with an NFW subhalo\footnote{The NFW subhalo is a localised perturber. For this parametric model fitting, we only use the potential correction results in the central region (indicated by the black circle in Figure~\ref{fig:sub_dpsi_only_kappa}), as the outskirts are likely degraded by noise.}, external shear, and an $m_4$ multipole perturber, and evaluate whether the fitting results agree with the true input values. For the Gaussian random field, we compare the power spectrum of the model-derived lensing potential perturbations with the true input values. As shown in Figure~\ref{fig:dpsi_inv_diagnosis}, the lensing potential perturbations derived by the potential correction method (second row) are consistent with the true input values (first row). A quantitative diagnostic (bottom row), conducted either via parametric model fitting or power spectrum comparison, demonstrates that our potential correction method can accurately recover the input perturbations. The specific parametric values from this diagnostic are summarised\footnote{It should be noted that, although the power spectrum properties of the input Gaussian random field are recovered reasonably well using the potential correction method, the model and the input true values do not perfectly align within the statistical uncertainty, a phenomenon also observed in \citet{Vernardos2022}.} in Table~\ref{tab:diagnose_inversion}.

\subsection{Simultaneous Inversion of Source and Potential Perturbations}
\label{sec:res_inv_dpsi_src}
\subsubsection{Initialisation with True Lens and Source Models}
\label{sec:res_inv_dpsi_src_true}

\begin{figure*}
	\includegraphics[width=\textwidth]{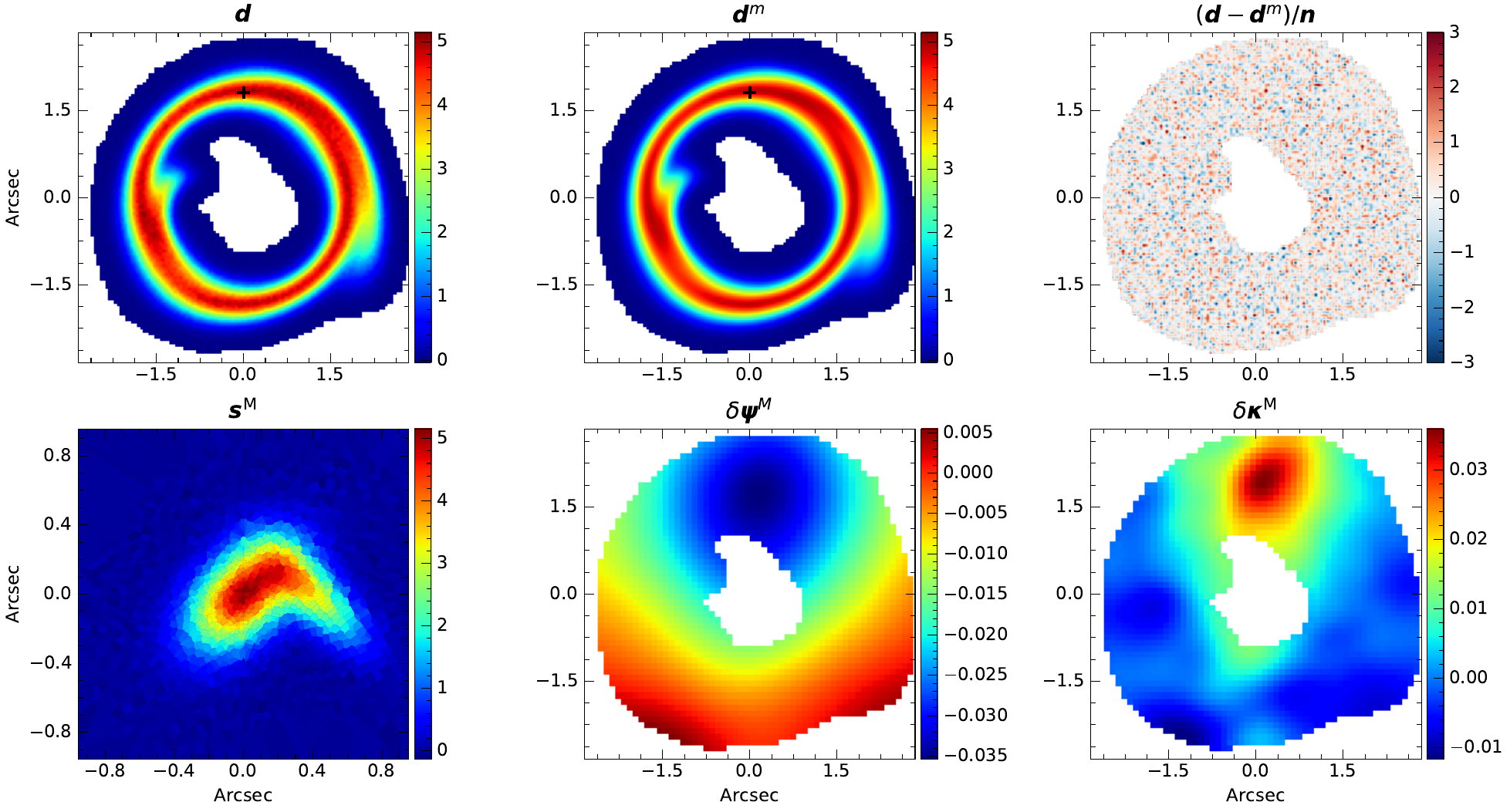}
        \caption{Results of simultaneous source and perturbative lensing potential modelling to directly fit the observed lensing image, using the input main lens mass and source light models as the starting point. From left to right: the first row displays the data ($\boldsymbol{d}$), model ($\boldsymbol{d}^m$), and normalised residual (($\boldsymbol{d} - \boldsymbol{d}^m)/\boldsymbol{n}$) images. The second row shows the reconstructed source ($\boldsymbol{s}^\mathrm{M}$), the lensing potential correction ($\delta \boldsymbol{\psi}^M$), and the corresponding convergence correction ($\delta \boldsymbol{\kappa}^\mathrm{M}$). The `plus' symbol in the top-left and top-middle panels marks the position of the input subhalo.}
    \label{fig:dpsi_src_inv_true_start_nfw}
\end{figure*}

\begin{figure*}
	\includegraphics[width=\textwidth]{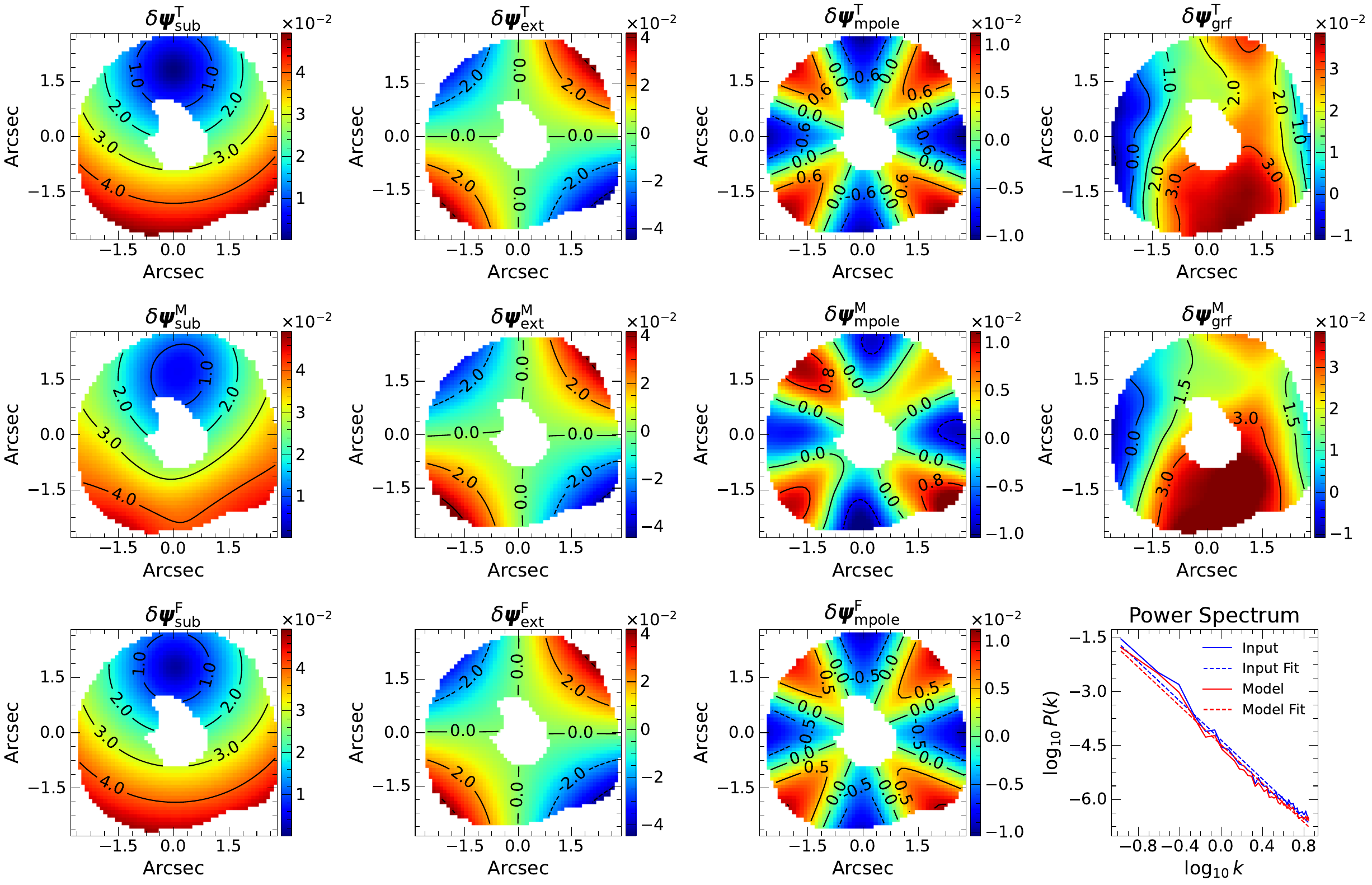}
        \caption{Analogous to Figure~\ref{fig:dpsi_inv_diagnosis}, but for the simultaneous modelling of the source and lensing potential perturbations, assuming perfect knowledge of the true main lens mass and source.}
    \label{fig:dpsi_src_inv_true_start_diagnosis}
\end{figure*}

We construct the matrices $\mathbf{L_s}$ and $\mathbf{L_{\delta \psi}}$, as defined in Equation~\eqref{eq:dpsi_src_linear_response}, using the true input models for the main lens mass and source light. The semi-linear inversion framework described in Section~\ref{sec:inv_dpsi_src} is then employed to solve for the pixelized source light ($\boldsymbol{s}$) and lensing potential perturbations ($\delta \boldsymbol{\psi}$). To regularise the solutions, \texttt{ConstantSplit} regularisation is applied to the pixelized source light, while Mat\'ern regularisation is used for the lensing potential perturbations. The hyperparameters associated with the source and lensing potential perturbation regularisations are determined by maximising the Bayesian evidence, and the specific values of the latter are presented in Table~\ref{tab:optimal_hyper_params}. 

As a demonstration, we present the results of simultaneously solving for the pixelized source and lensing potential perturbations for a simulated lens with an input NFW perturber (Figure~\ref{fig:dpsi_src_inv_true_start_nfw}). Our method effectively recovers the lensed image, reducing model image residuals to the noise level. The reconstructed source closely resembles the input. Furthermore, we clearly observe a localised subhalo signal in the perturbative lensing potential map (manifested as a potential well) and in the corresponding convergence map (manifested as a positive mass clump).

Figure~\ref{fig:dpsi_src_inv_true_start_diagnosis} compares the lensing potential perturbations derived from our method (second row) with the input truth (first row) for all four perturber types considered in this work: NFW subhalo, external shear, $m_4$ multipole, and Gaussian random field. The recovered perturbations agree closely with the input values. For a quantitative comparison, we fit each perturbation with a parametric model (for the NFW subhalo, external shear, and $m_4$ multipole) or calculate the power spectrum (for the Gaussian random field). We then verify whether the recovered perturbation properties, summarised in Table~\ref{tab:diagnose_inversion}, are consistent with the input truth. We find that the recovered properties are generally statistically consistent with the input. This performance is comparable to tests that solve solely for the lensing potential perturbations, demonstrating that the lens-source degeneracy does not significantly degrade the recovery of various lensing potential perturbations by our method.

\subsubsection{Initialisation with Macro Model Estimates}
\label{sec:res_inv_dpsi_src_macro}
A procedure similar to that described in Section~\ref{sec:res_inv_dpsi_src_true} is used to solve for the unknown source and lensing potential perturbations. However, the matrices $\mathbf{L_s}$ and $\mathbf{L_{\delta \psi}}$ are constructed using the macro-model results. As our macro-model employs a pixelized source model, we derive the source gradient as follows to construct $\mathbf{L_s}$. For each Voronoi source cell, a square cross is centred on the cell, aligned with the source plane's x and y axes. The side length of this square cross is equal to the square root of the cell's area. We then evaluate the source brightness at the four endpoints of the cross using natural neighbour interpolation. The source brightness gradient for that Voronoi cell is then calculated using a finite difference approximation between the endpoints. Once the source brightness gradient has been calculated for every Voronoi cell, the gradient at any position on the source plane can be determined via further natural neighbour interpolation.

\begin{figure*}
	\includegraphics[width=\textwidth]{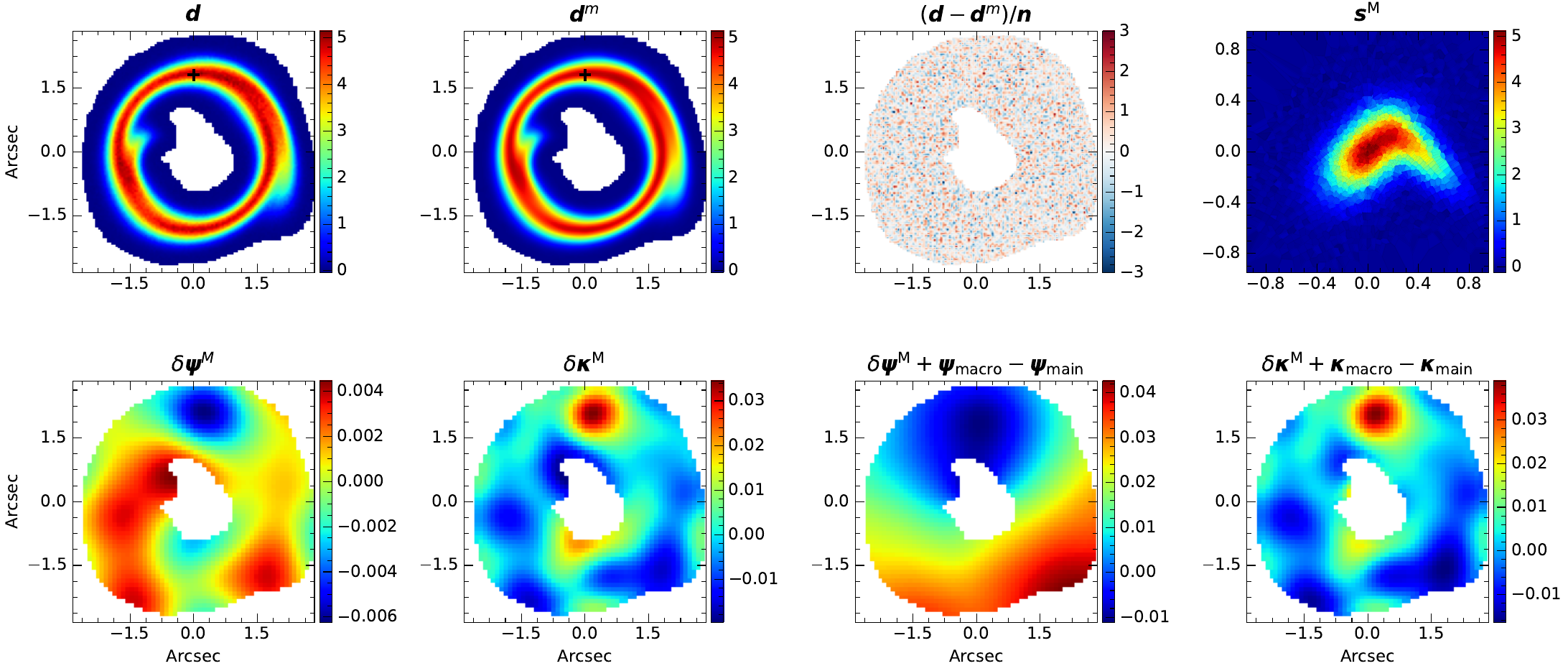}
        \caption{This figure replicates the structure of Figure~\ref{fig:dpsi_src_inv_true_start_nfw}, but initiates the potential correction from the main lens mass and source light models derived by the macro model. As the macro model yields a biased estimate of the main lens mass, we also present a ``pseudo-true'' lensing potential (convergence) perturbation map, defined as the model-derived perturbative lensing potential (convergence) plus the one given by the macro model minus the input true main lens values: $\delta \boldsymbol{\psi}^\mathrm{M} + \boldsymbol{\psi}_\mathrm{macro} - \boldsymbol{\psi}_\mathrm{main}$ $\left(\delta \boldsymbol{\kappa}^\mathrm{M} + \boldsymbol{\kappa}_\mathrm{macro} - \boldsymbol{\kappa}_\mathrm{main}\right)$.}
    \label{fig:dpsi_src_inv_nfw}
\end{figure*}

Figure~\ref{fig:dpsi_src_inv_nfw} presents the potential correction results for the mock lens with an input NFW subhalo perturber. The model continues to fit the data exceptionally well, with the remaining image residuals reduced to the noise level. A localised perturber, representing the signal from the input subhalo, is clearly visible in the potential correction (first column panel in the bottom row) and the corresponding convergence correction map (second column panel in the bottom row). Since the macro model can absorb some perturbation signals from the input subhalo and provides a biased estimate of the input true main lens mass and source, the perturbation reconstructed by the potential correction method does not constitute a direct comparison with the input true perturbation. Therefore, we define a ``pseudo-true perturbation'' quantity, which is the perturbation derived from the potential correction method, denoted as $\delta \boldsymbol{\psi}^\mathrm{M}$, plus the lensing potential predicted by the macro model, $\boldsymbol{\psi}_\mathrm{macro}$, minus the input true main lens values, $\boldsymbol{\psi}_\mathrm{main}$. This ``pseudo-true perturbation'' quantity is more comparable to the input true values of the perturbations.

\begin{figure*}
	\includegraphics[width=\textwidth]{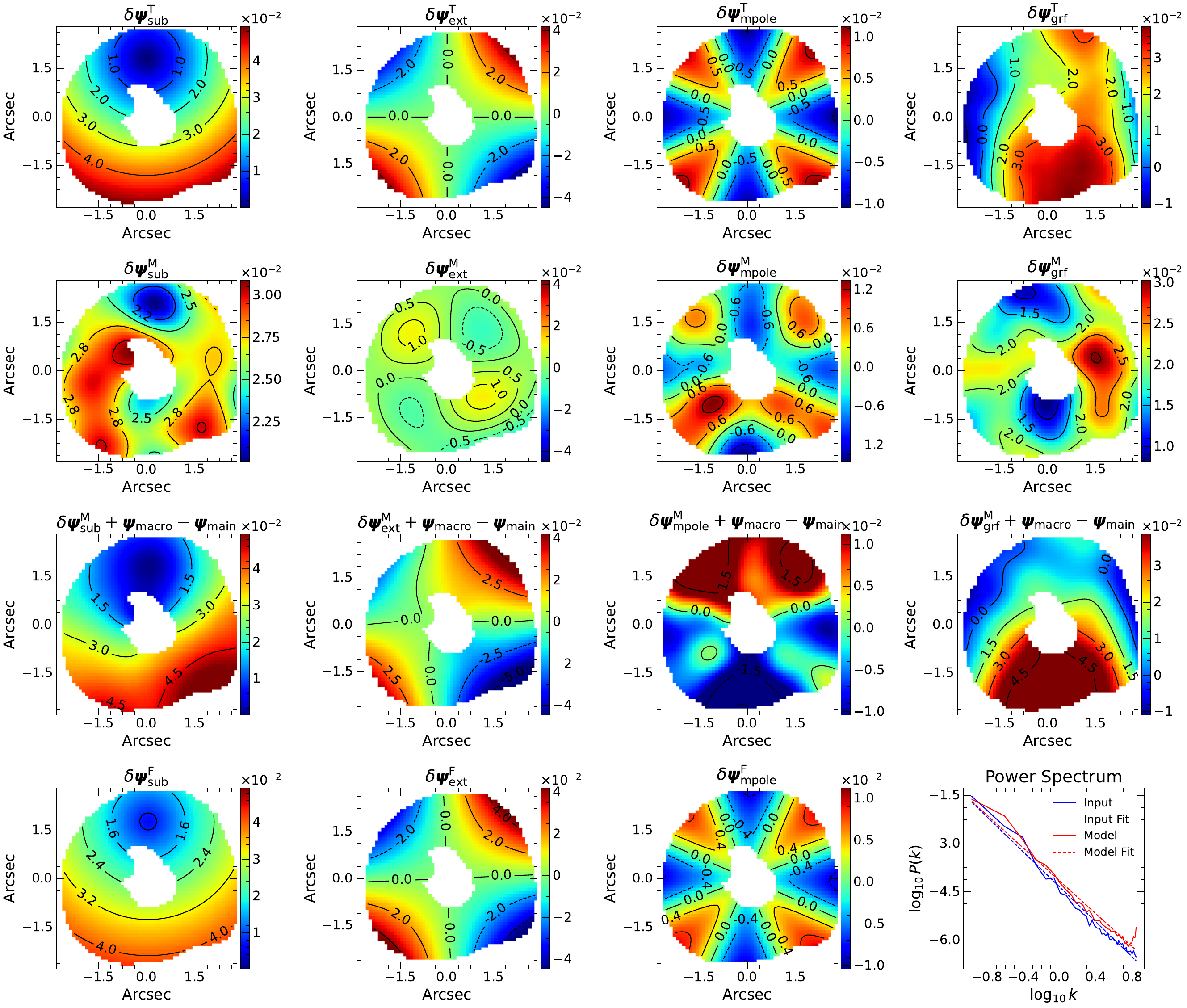}
        \caption{Analogous to Figure~\ref{fig:dpsi_src_inv_true_start_diagnosis}, but starting the potential correction with a biased estimate of the main lens mass and source light derived from the macro model. The third row displays the ``pseudo-true'' lensing potential perturbation maps, calculated as the sum of the model-derived perturbative lensing potential (convergence) and the macro model potential, minus the true input main lens potential ($\delta \boldsymbol{\psi}^\mathrm{M} + \boldsymbol{\psi}_\mathrm{macro} - \boldsymbol{\psi}_\mathrm{main}$). The quantitative diagnostics in the fourth column, including parametric model fitting and power spectrum calculation, are also based on these ``pseudo-true'' lensing potential perturbation maps.}
    \label{fig:dpsi_src_inv_diagnosis}
\end{figure*}

Figure~\ref{fig:dpsi_src_inv_diagnosis} is analogous to Figure~\ref{fig:dpsi_src_inv_true_start_diagnosis} and is employed to assess whether starting from a biased estimation of the main lens mass and source degrades the performance of the potential correction for all four types of perturbations examined in this work. It is noteworthy that while the localised subhalo and $m_4$ multipole signal remain directly visible in the perturbative lensing potential map derived from the potential correction method (second row), the model perturbations of shear and the Gaussian random field do not resemble the input at all. This does not indicate a failure of the potential correction method, as the ``pseudo-true perturbations'' for shear and the Gaussian random field (third row), which provide a more appropriate comparison, remain in good agreement with the input. However, initiating the potential correction with inaccurate estimates of the main lens mass and source introduces small biases in the recovered perturbation fields when conducting quantitative diagnoses. As shown in Table~\ref{tab:diagnose_inversion}, the shear orientation is offset by $\sim 2 \degr$ compared to the input truth, and the subhalo's mass exhibits a tendency to be underestimated by $\sim 0.5$ dex.

\subsection{Characterising the Subhalo in the Jackpot Lens via a Free-Form Approach}
\label{sec:res_0946}
To apply the potential correction method to the Jackpot lens, we first generate an annular mask around the lensed arc by connecting pixels with a lensed image SNR greater than 3 and expanding the connected region by 5 pixels. Next, we use the inferred mass model derived by \citet{He2024_mge}, where the main lens's mass is described by the EPL plus shear model, as the starting point to rerun a macro model that does not account for any perturbations. This refined macro model employs a pixelized source model with \texttt{Kmeans} pixelization and \texttt{AdaptiveBrightnessSplit} regularisation. The inferred mass model and the corresponding pixelized source reconstruction are then used to construct the $\mathbf{L}\left(\psi_p, s_p\right)$ matrix and solve for the unknown lensing potential perturbations, as detailed in Section~\ref{sec:inv_dpsi_src}. The hyperparameters associated with the source model, including \texttt{Kmeans} pixelization and \texttt{AdaptiveBrightnessSplit} regularisation, are fixed to the values obtained from the macro model when performing the potential correction. Allowing these hyperparameters to vary as free parameters is numerically unstable and renders the matrix equation~\ref{eq:dpsi_src_linear_response} unsolvable.

\begin{figure*}
	\includegraphics[width=\textwidth]{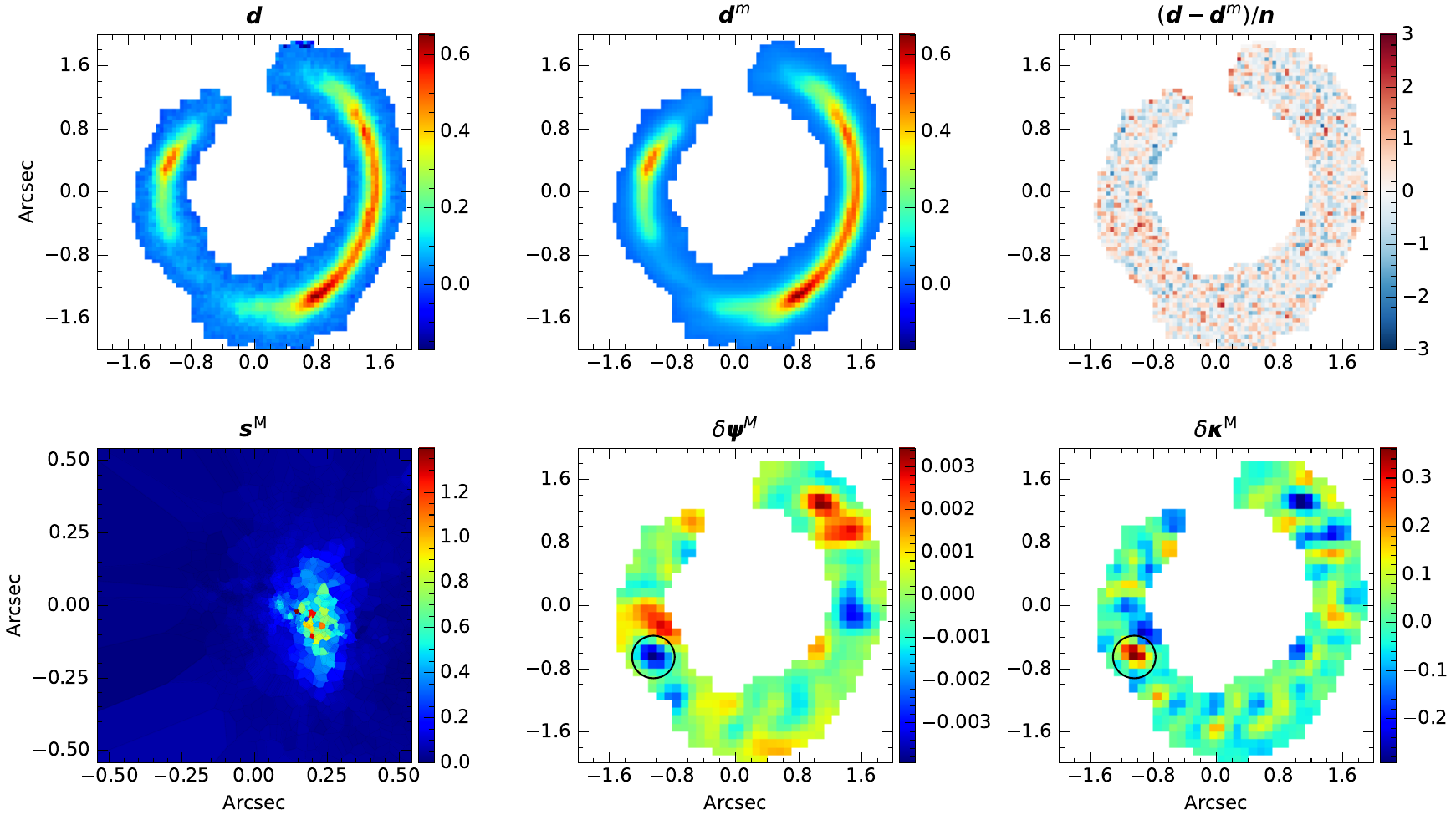}
        \caption{Potential corrections for SLACS0946+1006, analogous to Figure~\ref{fig:dpsi_src_inv_nfw}. In the bottom middle and right panels, black circles are centred on the peak of the localised convergence perturbation, with a radius of $0.27^{\prime\prime}$, corresponding to a physical scale of 1 kpc at the lens redshift.}
    \label{fig:dpsi_src_inv_0946}
\end{figure*}

Figure~\ref{fig:dpsi_src_inv_0946} presents the potential correction results for the Jackpot lens. Our model fits the observed lensing image exceptionally well, reducing the remaining image residuals to the noise level. A localised mass perturbation attributed to a subhalo is evident in the lensing potential perturbation map (manifested as a negative lensing potential well) and the corresponding convergence perturbation map (manifested as a positive mass clump). The central position of this localised mass perturber, defined as the pixel position with the maximum value in the convergence perturbation map, is $(y, x) = (-0.65, -1.05)$, consistent with the detection by \citet{Vegetti2010}. 

\citet{Minor2017_robust_radius} define a model-independent method to measure the ``robust mass'' of the subhalo, which is the aperture mass within the so-called robust radius $R_\text{robust}$. This radius can be calculated based on the geometric perturbations induced by the subhalo on the main lens's critical lines. For the subhalo in the Jackpot lens, $R_\text{robust} \approx 1$ kpc \citep{Minor2021_oc_subhalo}, as indicated by the black circle in Figure~\ref{fig:dpsi_src_inv_0946}. This yields a robust mass of $3 \times 10^8 M_{\odot}$ based on our potential correction result. However, this value is an order of magnitude lower than the result reported by \citet{Minor2021_oc_subhalo}, who explicitly used a truncated NFW model to describe the subhalo's mass, deriving a robust mass of $\sim 2-3.7 \times 10^9 M_{\odot}$ ($> \sim 95\%$ confidence level). We attribute this discrepancy to the limitations of our regularisation scheme, which, although it clearly performs better than the alternatives (as shown in Figure~\ref{fig:sub_dpsi_only_kappa}), still lacks the dynamic range needed to reconstruct highly compact perturbers. Specifically, we impose a constant regularisation strength for lensing potential perturbations across the entire modelling region. Although Bayesian evidence determines a relatively strong regularisation strength to suppress noise-like features in regions without significant perturbations, this approach simultaneously oversmooths the structure of highly compact perturbers. Notably, this limitation is analogous to the challenges faced when modelling clumpy sources with steep brightness gradients. In Section~\ref{sec:discuss_adpt_reg}, we discuss how adaptive regularisation for lensing potential perturbations could mitigate these systematics.

\begin{figure*}
	\includegraphics[width=\textwidth]{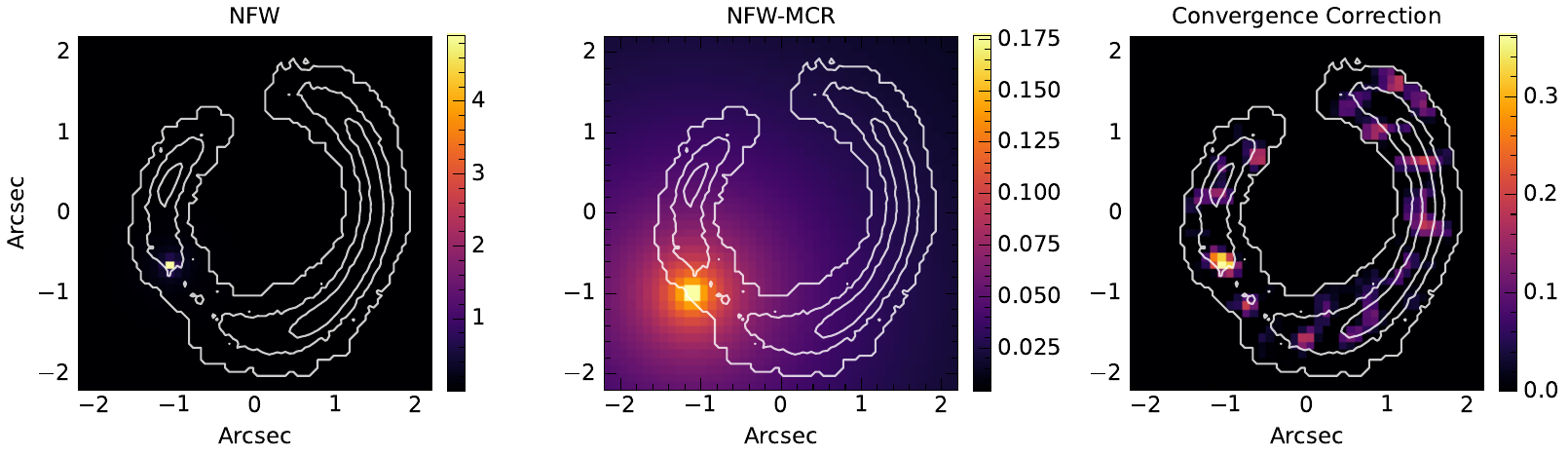}
        \caption{Convergence maps of the SLAC0946+1006 subhalo from different studies. \textbf{Left}: A $\sim 10^{10} \, M_{\odot}$ NFW subhalo with a high concentration of $\sim 200$, centered at $(-1.040, -0.651)$ \citep{Despali2024}. \textbf{Middle}: A $\sim 10^{11.5} \, M_{\odot}$ NFW subhalo following the standard mass-concentration relation, located at $(-1.00, -1.10)$ \citep{Nightingale2024}. \textbf{Right}: The perturbative convergence map resulting from our potential correction. Black contours indicate the lensed image morphology.}
    \label{fig:model_comparison_0946}
\end{figure*}

The properties of the localised perturber in the Jackpot lens have been systematically studied using parametric approaches. These studies assume the perturber is either a subhalo with various mass profiles, such as NFW, truncated NFW, NFW following the mass-concentration relation, pseudo-Jaffe, or power-law models \citep{Vegetti2010, Minor2021_oc_subhalo, Nightingale2024, Ballard2024_0946, Minor2024_0946, Despali2024}, or line-of-sight halos (LOS-halos) \citep{Enzi2024}. In the left and middle panels of Figure~\ref{fig:model_comparison_0946}, the convergence maps of the subhalo in the Jackpot lens, modelled by \citet{Despali2024} and \citet{Nightingale2024}, are compared, with the lensed arc shown as contours. The subhalo detected by \citet{Despali2024} exhibits a more concentrated structure and is positioned closer to the lensed arc compared to the modelling result of \citet{Nightingale2024}. This difference highlights how the assumed parametric functional form can influence subhalo measurements. We revisit this problem using a fully free-form approach that does not rely on any predefined functional form for the mass perturbations unaccounted for by the macro mass model. As shown in the right panel of Figure~\ref{fig:model_comparison_0946}, the mass perturber derived using our potential correction method exhibits a convergence distribution that strongly aligns with the parametric model result of \citet{Despali2024}, rather than that of \citet{Nightingale2024}, in terms of both position and compactness. This result represents the first application of a fully free-form model to reveal the compact nature of the subhalo in the Jackpot lens.

While the flexible nature of the free-form model allows it to avoid many systematics present in parametric approaches, this flexibility can also make the performance of free-form models less stable than parametric ones. Therefore, we conduct several tests to evaluate the robustness of our potential correction results for the Jackpot lens. First, we applied a larger annular mask to the Jackpot lens and performed the potential correction again, finding that the properties of the reconstructed subhalo remained unchanged. Second, the potential correction results reveal additional features in addition to the detected subhalo. For instance, a dipole-like structure appears near the detected subhalo (in the top-right region), and a negative convergence peak is observed near $(x, y) \approx (1.0, 1.0)$. To investigate the origin of these features, we employ a parametric approach analogous to that used by \citet{Despali2024} to model the Jackpot lens and simulate Jackpot-like mock lenses based on this model. Applying our potential correction method to these mock lenses reproduces the same additional features. Further analysis suggests that these arise from initiating the potential correction with a biased estimate of the main lens mass and source light provided by the macro model. Third, we simulate two Jackpot-like mock lenses with subhalo properties given by \citet{Despali2024} and \citet{Nightingale2024}, respectively. After applying our potential correction method to these mock lenses, we observe a localised perturber in both cases. Moreover, the perturber exhibits a more compact structure in the mock lens generated using the model of \citet{Despali2024}. This test demonstrates our potential correction method's ability to reveal the compactness of a localised perturber. A detailed description of the robustness tests conducted here is provided in Appendix~\ref{sec:appdx_B}.

\section{Discussion}
\label{sec:discuss}
Compared to prior works, our potential correction code offers several key advantages:
\begin{itemize}
    \item It simultaneously reconstructs the pixelized source and the lensing potential perturbation, thereby accounting for the covariance between them.
    \item It objectively determines the regularisation parameters for the pixelized source and the lensing potential perturbation by optimising the Bayesian evidence.
    \item Its versatility allows for the recovery of both localised and extended lensing potential perturbations, facilitated by Mat\'ern regularisation.
    \item The entire modelling procedure is automatic and requires no manual fine-tuning.
\end{itemize}
The efficacy of our programme has been demonstrated using both simulated and observational datasets, as presented in Section~\ref{sec:result}. Despite these impressive results, the current method has limitations, and there is scope for further improvement. These aspects are discussed in detail below.

\subsection{Enhanced Regularisation Approaches for Lensing Potential Perturbations}
\label{sec:discuss_adpt_reg}
When a pixelized free-form model is employed for lensing potential perturbations, strong regularisation is required to suppress noise-like signals in regions where perturbations are absent. Conversely, in regions where localised perturbations with high spatial gradients emerge, the regularisation strength should be reduced to allow more perturbation structures to be determined from the data rather than being overly smoothed by the prior (i.e., the regularisation). Assuming a constant regularisation strength across the entire modelling region fails to satisfy the differing requirements of regions with and without perturbations simultaneously. The optimal regularisation strength, as determined by Bayesian evidence, must serve as a compromise: it will be too weak for regions lacking perturbations and too strong for regions with highly localised perturbations. Consequently, a constant regularisation scheme tends to over-smooth the structures of highly localised perturbations, such as subhalos. To fully address these systematic issues, adaptive regularisation schemes that allow for a broader dynamic range in lensing potential perturbations can be employed. Specifically, when localised perturbations are identified in the potential correction results obtained through a constant regularisation scheme, this information can be utilised as a prior. The regularisation strength can then be adjusted accordingly, increasing it in regions lacking significant mass signals and decreasing it in areas where localised mass structures are present.

Our potential correction framework is based on the quadratic form of the regularisation function, which allows the unknown perturbations in the lensing potential to be solved linearly. It has been demonstrated that quadratic regularisation with the Mat\'ern kernel can effectively recover various types of lensing potential perturbations, whether extended or localised. Despite these impressive results, we observe certain artefacts in the potential correction results, such as random negative-convergence features, which have also been noted in previous works \citep{Vegetti2010}. These artefacts stem from the inherent nature of quadratic regularisation (also known as L2 regularisation or ridge regression), which promotes smoothness by shrinking all feature values (specifically, the lensing potential perturbations at each pixel) without setting any to zero \citep{L2_reg_1,L2_reg_2}. To improve the potential correction method, an alternative approach to imposing smoothness priors may involve setting the lensing potential perturbation values at each pixel to zero, unless the data indicate otherwise. This requirement aligns with the purpose of L1 regularisation \citep[or LASSO;][]{L1_reg}. However, utilising L1 regularisation disrupts the linear inversion framework, preventing us from using equation~\ref{eq:dpsi_only_mp_inv} to solve for the lensing potential perturbation. Instead, we must determine the lensing potential perturbation values by directly optimising the penalty function illustrated in equation~\ref{eq:dpsi_only_full_panalty}. This approach introduces a substantial number of nonlinear modelling parameters (several thousand), which are challenging for optimisers that do not leverage gradient information to navigate parameter space. Since our current code lacks support for auto-differentiation, we are unable to compute the gradient of the likelihood with respect to the modelling parameters, rendering our code incompatible with L1 regularisation at this time. Several open-source lens modelling software packages have incorporated auto-differentiation \citep{Gu2022, Galan2022_wavelet, Cao2025_jax}, and \texttt{PyAutoLens} plans to integrate this functionality soon. We plan to incorporate L1 regularisation (and potentially more general forms) into our code in future work.

\subsection{Quantifying Uncertainties in Lensing Potential Perturbations}
\label{sec:discuss_error}
The artefacts in our current potential correction results are unavoidable due to the limitations of quadratic regularisation schemes. However, if we can quantify the uncertainties in the model-derived lensing potential perturbations and demonstrate that the signal in these artefact regions is highly uncertain, these artefacts will not impede the interpretation of the potential correction results. Given a set of hyperparameters for the source and lensing potential perturbation regularisations, which are best determined by maximizing the Bayesian evidence, we can estimate the uncertainties of the pixelized source and lensing potential perturbations using the covariance matrix, the mathematical form of which is provided in Equation~\ref{eq:dpsi_only_cov_mat}\footnote{Replace $\mathbf{L}_{\delta \psi}$ with $\mathbf{L}$.}. This covariance matrix contains numerous non-zero off-diagonal elements, indicating a high degree of correlation between the pixelized source and lensing potential perturbation reconstructions. Since our primary interest lies in determining whether a mass perturbation structure is genuine, we aim to quantify the uncertainties of pixelized mass perturbations and define an SNR-like quantity to evaluate the fidelity of the perturbation signal. However, we currently lack a robust method to translate the covariance matrix of the lensing potential perturbations into uncertainties for the mass perturbations. Consequently, we resort to Monte Carlo simulations to estimate the uncertainties of pixelized mass perturbations, as detailed in Appendix~\ref{sec:appdx_B}.

\subsection{Incorporating Lens Light Modelling within Potential Correction Algorithms}
\label{sec:discuss_lens_light}
The framework presented in Section~\ref{sec:inv_dpsi_src} implicitly assumes that the light contribution from the lens galaxy has been perfectly subtracted from the lensing image prior to applying the potential correction method. However, pre-subtracting lens light limits our ability to fully account for the covariance between lens light and lensed source light. It has been demonstrated that lensing potential perturbations can be partially degenerate with the lens light signal \citep{Nightingale2024}. Therefore, a more effective approach to managing lens light is to integrate its modelling within the potential correction framework. This can be achieved using the mathematical framework presented in \citet{He2024_mge, Ritondale2019}. 

Suppose we utilise $N_l$ elliptical Gaussian models that share a common centre and ellipticity, but have different standard deviations equally spaced in a logarithmic radial range (e.g.\ from $10^{-2}$ to 3$\arcsec$) to represent the lens light. The image $\boldsymbol{d^l}$ formed by these Gaussian models can be expressed in the following matrix equation:
\begin{equation}
\boldsymbol{d^l} = \boldsymbol{B} \mathbf{L_l}(\xi_l) \boldsymbol{l}.
\label{eq:lens_light_response}
\end{equation}
Here, the matrix $\mathbf{L_l}(\xi_l)$ has dimensions $[N_d, N_l]$, with each column representing the model image formed by each Gaussian model with a "unit" normalization value; $\xi_l$ denotes the nonlinear parameters corresponding to the centre and ellipticities of the Gaussian models in this instance. The column vector $\boldsymbol{l}$ contains $N_l$ elements, representing the normalisation factors for each Gaussian model. Equation~\ref{eq:dpsi_src_linear_response} can now be extended to include lens light modelling as follows:
\begin{equation}
\begin{aligned}
\boldsymbol{d} &= \boldsymbol{B} \mathbf{L_s}(\boldsymbol{\psi}_p) \boldsymbol{s} + \boldsymbol{B} \mathbf{L_{\delta \psi}}(\boldsymbol{s}_p) \boldsymbol{\delta \psi} + \boldsymbol{B} \mathbf{L_l}(\xi_l) \boldsymbol{l} + \boldsymbol{n} \\
&= \boldsymbol{B} \mathbf{L}_\text{all}(\boldsymbol{\psi}_p, \boldsymbol{s}_p, \xi_l) \boldsymbol{r}_\text{all} + \boldsymbol{n}.
\end{aligned}
\label{eq:dpsi_src_lens_light_linear_response}
\end{equation}
Here, the block matrices
\begin{equation}
\mathbf{L}_\text{all}(\boldsymbol{\psi}_p, \boldsymbol{s}_p, \xi_l) \equiv \left(\mathbf{L_s}(\boldsymbol{\psi}_p) \mid \mathbf{L_{\delta \psi}}(\boldsymbol{s}_p) \mid \mathbf{L_l}(\xi_l)\right),
\end{equation}
and
\begin{equation} 
\boldsymbol{r}_\text{all} \equiv \left(\begin{array}{c} \boldsymbol{s} \\ \delta \boldsymbol{\psi} \\ \boldsymbol{l} \end{array}\right).
\end{equation}
The Hessian of the regularisation function for the vector $\boldsymbol{r}_\text{all}$ can be expressed as:
\begin{equation}
\mathbf{R}_{r_\text{all}}=
\begin{pmatrix}
\mathbf{R}_s & \mathbf{0} & \mathbf{0}\\
\mathbf{0} & \mathbf{R}_{\delta\psi} & \mathbf{0} \\
\mathbf{0} & \mathbf{0} & \epsilon\mathbf{I}_l
\end{pmatrix},
\end{equation}
where $\mathbf{I}_l$ is an identity matrix of dimensions $[N_l, N_l]$, and $\epsilon$ is a small constant (e.g., $10^{-5}$). Equations~\ref{eq:dpsi_src_linear_response} and \ref{eq:dpsi_src_lens_light_linear_response} retain the same mathematical form; thus, the linear inversion framework outlined in Section~\ref{sec:inv_dpsi_src} remains applicable for potential corrections that incorporate lens light modelling.

\subsection{Concurrent Modelling of Main Lens and Lensing Potential Perturbations}
\label{sec:discuss_main_lens_pert_model_together}
In Section~\ref{sec:res_inv_dpsi_src_macro}, we demonstrated that the perturbations derived using the potential correction method do not generally match the input ground truth, as the macro model can vary to absorb some of the mass perturbations. To enable the potential correction method to recover the underlying true perturbations, a natural approach would be to simultaneously fit the main lens mass and the lensing potential perturbations \citep[e.g.][]{Vernardos2022}. Although we cannot currently perform this simultaneous fitting due to a fixed main lens mass model being hard-coded in our programme, we observe that starting the potential correction from the input true main lens mass model yields higher Bayesian evidence than starting from the macro model's estimate. This suggests that simultaneous fitting of the main lens mass and lensing potential perturbations might be feasible. We plan to explore this possibility in future work.

\section{Summary and conclusion}
\label{sec:summary}
We introduce a new free-form lens modelling tool built upon the open-source software \texttt{PyAutoLens}. Our approach models perturbations to the smooth lensing potential derived by a macro model, as free parameters defined on a pixelized grid. By applying only minimal smoothing priors to these pixelized perturbations, we mitigate many systematic errors commonly encountered in parametric lens modelling arising from the strong priors imposed by predefined functional forms. Our key results are:

\begin{itemize}
    \item Our code can recover various forms of perturbation signals, both extended and localised, in an automated manner. It avoids the fine-tuning issues that plague iterative potential correction methods by employing the Mat\'ern regularisation kernel and objectively determining the hyperparameters for lensing potential perturbations and source regularisations through Bayesian evidence optimisation. Although initially developed for substructure detection, our code's ability to recover various forms of mass perturbation can also benefit other applications requiring the assessment of biases introduced by assumed parametric functional forms, such as time-delay cosmography \citep[e.g.][]{Suyu2009_pt, Kochanek2021}.
    
    \item Properties of the lensing potential perturbations are recovered without significant bias in mock tests. Objectively determining the regularisation parameters enhances the capability of our potential correction method to quantitatively reproduce the perturbation field. 
    
    \item Degeneracies between the lens and source do not significantly impair the ability of our correction method to recover the properties of lensing potential perturbations, even when starting from a biased initial estimate of the lens and source, although there is a slight reduction in accuracy.
    
    \item Our analysis reveals a localised mass concentration within the Jackpot strong-lensing system. Its position and compact morphology are consistent with the subhalo detection obtained from parametric modelling by \citet{Vegetti2010}, but contrast with the results of \citet{Nightingale2024}. Because our method is entirely free-form, it circumvents several systematic uncertainties inherent to parametric lens modelling, particularly the degeneracy between the subhalo and the main lens mass distributions. Consequently, our findings provide compelling evidence that the subhalo in the Jackpot strong lens is over-concentrated, potentially challenging the standard cosmological model \citep[e.g.][]{Minor2021_oc_subhalo}. These results may necessitate alternative self-interacting dark matter models to explain the anomalous subhalo properties observed in the Jackpot lens \citep[e.g.][]{Shubo2025}.
\end{itemize}

Our current implementation applies constant regularisation across the entire modelling area. While this constant regularisation effectively suppresses noise in regions without obvious perturbations, it overly smooths the structure of highly compact mass clumps, particularly within the central cuspy region. Consequently, our method tends to underestimate the mass of highly compact mass clumps, such as a $10^{10} M_{\odot}$ NFW subhalo with a high concentration of $\sim$200. Future work will involve implementing adaptive regularisation to overcome this limitation. We anticipate incorporating an auto-differentiation framework into our potential correction method. This enhancement would provide greater flexibility in selecting alternative regularisation forms (beyond the quadratic form) and offer a more robust framework for estimating uncertainties or the significance of features reconstructed by the potential correction method.

This work represents ongoing progress in the development of free-form lens modelling tools. Over the coming years, large-scale surveys such as Euclid, CSST, and Roman are anticipated to observe hundreds of thousands of galaxy–galaxy strong lenses \citep{Collett2015, Cao2024, Weiner2020, Nagam2025}. These surveys may yield thousands of high-quality strong-lensing systems suitable for subhalo detection \citep{Riordan2023}, opening a new observational window into the properties of dwarf satellite galaxies beyond the Local Universe \citep{Kaihao2025}. The objective and automated characteristics of our potential correction code render it especially suitable for analysing such extensive lens samples.

\section*{Acknowledgements}
We thank the referee for the helpful comments and suggestions that have improved this paper. This work was supported by the National Key R\&D Program of China (grant number 2022YFF0503403), the National Natural Science Foundation of China (No. 11988101), the K.C.Wong Education Foundation, the Ministry of Science and Technology of China (No. 2020SKA0110100), and the science research grants from China Manned Space Project with Nos.CMS-CSST-2021-B01 and CMS-CSST-2025-A03. XYC acknowledges the support of the National Natural Science Foundation of China (No.\ 12303006). RL is also supported by the National Natural Science Foundation of China (Nos 11773032, 12022306), the CAS Project for Young Scientists in Basic Research (No. YSBR-062). This work was supported in the UK by STFC via grant ST/X001075/1, the UK Space Agency via grant ST/W002612/1, and the European Research Council via grant DMIDAS (GA 786910). We express our gratitude to ChatGPT, an AI model developed by OpenAI, for its assistance in polishing the English of this paper.

\section*{Data Availability}
The code and data product that supports this work are publicly available from \url{https://github.com/caoxiaoyue/lensing_potential_correction}.



\bibliographystyle{mnras}
\bibliography{reference} 



\appendix
\section{Bayesian Framework for Potential Correction}
\label{sec:appdx_C}
Equation~\ref{eq:dpsi_linear_response} shows that $\delta \boldsymbol{d}$ and $\boldsymbol{\delta \psi}$ are connected via a linear mapping. The most likely solution for $\boldsymbol{\delta \psi}$ can be derived by maximizing the likelihood
\begin{equation}
P(\boldsymbol{\delta d} \mid \boldsymbol{\delta \psi}, \mathbf{L_{\delta \psi}}) = \frac{\exp\left(-E_{\mathrm{D}}(\boldsymbol{\delta d} \mid \boldsymbol{\delta \psi}, \mathbf{L_{\delta \psi}})\right)}{Z_{\mathrm{D}}},
\label{eq:dpsi_only_like}
\end{equation}
where
\begin{equation}
\begin{aligned}
E_{\mathrm{D}}(\boldsymbol{\delta d} \mid \boldsymbol{\delta \psi}, \mathbf{L_{\delta \psi}}) &= \frac{1}{2} (\mathbf{L_{\delta \psi}} \boldsymbol{\delta \psi} - \boldsymbol{\delta d})^{\mathrm{T}} \mathrm{C}_{\mathrm{D}}^{-1} (\mathbf{L_{\delta \psi}} \boldsymbol{\delta \psi} - \boldsymbol{\delta d}) \\
&= \frac{1}{2} \chi^2,
\end{aligned}
\label{eq:dpsi_only_ED}
\end{equation}
and
\begin{equation}
Z_{\mathrm{D}} = (2 \pi)^{N_{\mathrm{d}} / 2} \left( \operatorname{det} \mathbf{C}_{\mathrm{D}} \right)^{1 / 2}.
\label{eq:dpsi_only_ZD}
\end{equation}
This is equivalent to minimizing $E_{\mathrm{D}}$ with a linear least squares algorithm, which yields
\begin{equation}
\boldsymbol{\delta \psi}_{\mathrm{ml}} = \left( \mathbf{L^T_{\delta \psi}} \mathrm{C}_{\mathrm{D}}^{-1} \mathbf{L_{\delta \psi}} \right)^{-1} \mathbf{L^T_{\delta \psi}} \mathrm{C}_{\mathrm{D}}^{-1} \boldsymbol{\delta d}.
\label{eq:dpsi_only_ml_inv}
\end{equation}
The most likely solution $\boldsymbol{\delta \psi}_{\mathrm{ml}}$ is usually degraded by noise and is ill-posed \citep{Warren2003}. An additional prior $P(\delta \boldsymbol{\psi} \mid \boldsymbol{g}_{\delta \psi}, \boldsymbol{\xi}_{\delta \psi})$ is needed to impose smoothness on $\delta \boldsymbol{\psi}$, a process commonly called ``regularisation''. Here, $\boldsymbol{g}_{\delta \psi}$ represents the type of regularisation, and the associated model parameters are denoted by the vector $\boldsymbol{\xi}_{\delta \psi}$. The prior can be expressed as:
\begin{equation}
P(\delta \boldsymbol{\psi} \mid \boldsymbol{g}_{\delta \psi}, \boldsymbol{\xi}_{\delta \psi}) = \frac{\exp \left( -E_{\mathrm{\delta \psi}}(\delta \boldsymbol{\psi} \mid \boldsymbol{g}_{\delta \psi}, \boldsymbol{\xi}_{\delta \psi}) \right)}{Z_{\delta \psi}(\boldsymbol{\xi}_{\delta \psi})},
\label{eq:dpsi_only_prior}
\end{equation}
where $Z_{\delta \psi}(\boldsymbol{\xi}_{\delta \psi}) = \int \mathrm{d}^{N_p}(\delta \boldsymbol{\psi}) \exp(-E_{\mathrm{\delta \psi}})$ is the normalization factor of the prior probability density function. The function $E_{\delta \psi}$, also known as the regularisation function, typically takes a quadratic form: $E_{\delta \psi}(\delta \boldsymbol{\psi}) = \frac{1}{2} \delta \boldsymbol{\psi}^{\mathrm{T}} \mathbf{R_{\delta \psi}} \delta \boldsymbol{\psi}$, where $\mathbf{R_{\delta \psi}}$ represents the Hessian of $E_{\delta \psi}$. This form allows $\delta \boldsymbol{\psi}$ to be directly solved through matrix inversion \citep{Warren2003}. For more detailed information on the specific form of the regularisation function, refer to Section~\ref{sec:reg_func}.

In Bayesian inference, the posterior distribution of $\delta \boldsymbol{\psi}$ is obtained by multiplying the likelihood function by the regularisation prior, as follows:
\begin{equation}
P(\delta \boldsymbol{\psi} \mid \delta \boldsymbol{d}, \mathbf{L_{\delta \psi}}, \boldsymbol{\xi}_{\delta \psi}, \boldsymbol{g}_{\delta \psi}) = \frac{P(\delta \boldsymbol{d} \mid \delta \boldsymbol{\psi}, \mathbf{L_{\delta \psi}}) P(\delta \boldsymbol{\psi} \mid \boldsymbol{\xi}_{\delta \psi}, \boldsymbol{g}_{\delta \psi})}{P(\delta \boldsymbol{d} \mid \mathbf{L_{\delta \psi}}, \boldsymbol{\xi}_{\delta \psi}, \boldsymbol{g}_{\delta \psi})}.
\label{eq:dpsi_only_posterior}
\end{equation}
Here, the normalization factor $P(\delta \boldsymbol{d} \mid \mathbf{L_{\delta \psi}}, \boldsymbol{\xi}_{\delta \psi}, \boldsymbol{g}_{\delta \psi})$, known as Bayesian evidence, can be used to determine the choice of $\{\mathbf{L_{\delta \psi}}, \boldsymbol{g}_{\delta \psi}, \boldsymbol{\xi}_{\delta \psi}\}$ \citep{Suyu2006_bayes, Suyu2009_pt}. The most probable solution of $\delta \boldsymbol{\psi}$ maximizes $P(\delta \boldsymbol{\psi} \mid \delta \boldsymbol{d}, \mathbf{L_{\delta \psi}}, \boldsymbol{\xi}_{\delta \psi}, \boldsymbol{g}_{\delta \psi})$, which is equivalent to minimizing the following penalty function:
\begin{equation}
M(\delta \boldsymbol{\psi}) = E_{\mathrm{D}}(\delta \boldsymbol{\psi}) + E_{\delta \psi}(\delta \boldsymbol{\psi}).
\label{eq:dpsi_only_full_panalty}
\end{equation}
This yields Equation~\ref{eq:dpsi_only_mp_inv}:
\[
\boldsymbol{\delta \psi}_{\mathrm{MAP}} \;=\; \left( \mathbf{L}^{\mathrm{T}}_{\delta \psi}\, \mathbf{C}_{\mathrm{D}}^{-1}\, \mathbf{L}_{\delta \psi} + \mathbf{R}_{\delta \psi} \right)^{-1} \mathbf{L}^{\mathrm{T}}_{\delta \psi}\, \mathbf{C}_{\mathrm{D}}^{-1}\, \delta \boldsymbol{d},
\]
which indicates that, given the macro-model outputs $\{\delta \boldsymbol{d},\, \mathbf{L}_{\delta \psi}\}$ and an arbitrary choice of regularisation $\mathbf{R}_{\delta \psi}(\boldsymbol{\xi}_{\delta \psi}, \boldsymbol{g}_{\delta \psi})$, the most probable solution, $\boldsymbol{\delta \psi}_{\mathrm{MAP}}$, can be obtained via linear matrix inversion. The covariance matrix ($\Sigma_{\delta \psi}$), which describes the uncertainty of the most probable solution ($\boldsymbol{\delta \psi}_{\mathrm{mp}}$), is derived by \citet{Suyu2006_bayes} and follows:
\begin{equation}
\Sigma_{\delta \psi} = \left(\mathbf{L}_{\delta \psi}^T \mathrm{C}_\mathrm{D}^{-1} \mathbf{L}_{\delta \psi} + \mathbf{R}_{\delta \psi}\right)^{-1}.
\label{eq:dpsi_only_cov_mat}
\end{equation}

For a given choice of regularisation type $\boldsymbol{g}_{\delta \psi}$, the values of its hyperparameters $\boldsymbol{\xi}_{\delta \psi}$ can be optimally determined by maximizing the posterior $P(\boldsymbol{\xi}_{\delta \psi} \mid \delta \boldsymbol{d}, \mathbf{L}_{\delta \psi}, \boldsymbol{g}_{\delta \psi})$, where
\begin{equation}
P(\boldsymbol{\xi}_{\delta \psi} \mid \delta \boldsymbol{d}, \mathbf{L}_{\delta \psi}, \boldsymbol{g}_{\delta \psi}) = \frac{P(\delta \boldsymbol{d} \mid \mathbf{L}_{\delta \psi}, \boldsymbol{\xi}_{\delta \psi}, \boldsymbol{g}_{\delta \psi}) P(\boldsymbol{\xi}_{\delta \psi})}{P(\delta \boldsymbol{d} \mid \mathbf{L}_{\delta \psi}, \boldsymbol{g}_{\delta \psi})}.
\label{eq:dpsi_only_reg_par_posterior}
\end{equation}
Here, the evidence term $P(\delta \boldsymbol{d} \mid \mathbf{L}_{\delta \psi}, \boldsymbol{g}_{\delta \psi})$ marginalizes over the hyperparameters of the regularisation and can be used to rank different model choices of $\{\mathbf{L}_{\delta \psi}, \boldsymbol{g}_{\delta \psi}\}$. Since a non-informative prior is usually taken for $P(\boldsymbol{\xi}_{\delta \psi})$, finding the most probable solution for $\boldsymbol{\xi}_{\delta \psi}$ is equivalent to maximizing $P(\delta \boldsymbol{d} \mid \mathbf{L_{\delta \psi}}, \boldsymbol{\xi}_{\delta \psi}, \boldsymbol{g}_{\delta \psi}) \equiv \mathcal{E}$, whose explicit form is given in Equation~\ref{eq:dpsi_only_ev_eq}.
\section{Optimal Hyperparameter Values for the Matern Kernel}
\label{sec:appdx_A}
In this work, the Matérn kernel is employed to regularise the pixelised potential corrections. As defined in Equation~\ref{eq:matern_kernel}, the Matérn kernel is characterised by three hyperparameters: a coefficient $\lambda^{\delta \psi}$ that controls the overall smoothing strength; a characteristic length-scale $\rho$ (in units of arcseconds) that governs the smoothing scale; and a parameter $\nu$ that determines the order of the smoothing, ensuring the reconstructed field is $\left\lceil \nu \right\rceil - 1$ times differentiable. We assign non-informative priors to these three hyperparameters: $\lambda^{\delta \psi} \sim \mathcal{L}(10^{-6}, 10^6)$, $\rho \sim \mathcal{L}(10^{-4}, 10^3)$, and $\nu \sim \mathcal{U}(0.5, 10.0)$. The optimal values for these hyperparameters, presented in Table~\ref{tab:optimal_hyper_params}, were determined by sampling the Bayesian evidence.

\begin{table*}
\renewcommand{\arraystretch}{1.5}
\begin{tabular}{|ll|l|l|l|}
\hline
\multicolumn{2}{|l|}{} & $\log[\texttt{Coefficient}]$ & $\log[\texttt{Scale}]$ & \texttt{Nu} \\ \hline
\multicolumn{1}{|l|}{\multirow{3}{*}{\textbf{Spherical NFW Subhalo}}} & \textbf{$\delta \boldsymbol{\psi}$-only} & $3.1919^{+0.3860}_{-0.5423}$ & $0.6997^{+0.1881}_{-0.1426}$ & $2.2389^{+0.3510}_{-0.2562}$ \\
\multicolumn{1}{|l|}{} & \textbf{$(\delta \boldsymbol{\psi}, \boldsymbol{s})^\mathrm{Truth}$} & $3.5091^{+0.4556}_{-0.6482}$ & $0.5136^{+0.2297}_{-0.1791}$ & $2.8750^{+1.9273}_{-0.6309}$ \\
\multicolumn{1}{|l|}{} & \textbf{$(\delta \boldsymbol{\psi}, \boldsymbol{s})^\mathrm{Macro}$} & $4.9545^{+0.2624}_{-0.3180}$ & $0.0182^{+0.0877}_{-0.0737}$ & $6.5831^{+2.3161}_{-2.3562}$ \\ \hline
\multicolumn{1}{|l|}{\multirow{3}{*}{\textbf{Shear}}} & \textbf{$\delta \boldsymbol{\psi}$-only} & $2.0746^{+0.1259}_{-0.0556}$ & $0.7473^{+0.0712}_{-0.0578}$ & $5.5018^{+3.0553}_{-2.1787}$ \\
\multicolumn{1}{|l|}{} & \textbf{$(\delta \boldsymbol{\psi}, \boldsymbol{s})^\mathrm{Truth}$} & $2.0900^{+0.1322}_{-0.0662}$ & $0.7020^{+0.0628}_{-0.0590}$ & $7.2722^{+1.8435}_{-2.0310}$ \\
\multicolumn{1}{|l|}{} & \textbf{$(\delta \boldsymbol{\psi}, \boldsymbol{s})^\mathrm{Macro}$} & $3.5641^{+0.2972}_{-0.3101}$ & $0.2702^{+0.0547}_{-0.0551}$ & $8.3728^{+1.1427}_{-1.8146}$ \\ \hline
\multicolumn{1}{|l|}{\multirow{3}{*}{\textbf{Multipole}}} & \textbf{$\delta \boldsymbol{\psi}$-only} & $4.3240^{+0.1558}_{-0.1930}$ & $0.0260^{+0.0733}_{-0.0523}$ & $5.6456^{+2.2337}_{-1.4727}$ \\
\multicolumn{1}{|l|}{} & \textbf{$(\delta \boldsymbol{\psi}, \boldsymbol{s})^\mathrm{Truth}$} & $4.1447^{+0.2028}_{-0.2470}$ & $0.0518^{+0.0845}_{-0.0572}$ & $6.1994^{+2.3352}_{-1.9377}$ \\
\multicolumn{1}{|l|}{} & \textbf{$(\delta \boldsymbol{\psi}, \boldsymbol{s})^\mathrm{Macro}$} & $4.2074^{+0.1840}_{-0.2042}$ & $-0.0091^{+0.0805}_{-0.0542}$ & $6.1800^{+2.4526}_{-2.0224}$ \\ \hline
\multicolumn{1}{|l|}{\multirow{3}{*}{\textbf{Gaussian Random Field}}} & \textbf{$\delta \boldsymbol{\psi}$-only} & $2.6879^{+0.4094}_{-0.4496}$ & $0.7515^{+0.1646}_{-0.1663}$ & $1.7675^{+0.1484}_{-0.1261}$ \\
\multicolumn{1}{|l|}{} & \textbf{$(\delta \boldsymbol{\psi}, \boldsymbol{s})^\mathrm{Truth}$} & $2.8467^{+0.4500}_{-0.5086}$ & $0.6596^{+0.1992}_{-0.1939}$ & $1.8761^{+0.2691}_{-0.1987}$ \\
\multicolumn{1}{|l|}{} & \textbf{$(\delta \boldsymbol{\psi}, \boldsymbol{s})^\mathrm{Macro}$} & $4.4075^{+0.2332}_{-0.3383}$ & $0.0351^{+0.1423}_{-0.0869}$ & $3.8699^{+2.8851}_{-1.2515}$ \\ \hline
\end{tabular}
\caption{Optimal hyperparameter values for regularising lensing potential perturbations using the Mat\'ern kernel, determined objectively by sampling the Bayesian evidence. Results are shown for inversions of the lensing potential only ($\delta \boldsymbol{\psi}$-only), joint inversions of the source and lensing potential using the true main lens and source models ($(\delta \boldsymbol{\psi}, \boldsymbol{s})^\mathrm{Truth}$), and joint inversions initialised from the point estimate of the macro model without potential perturbations ($(\delta \boldsymbol{\psi}, \boldsymbol{s})^\mathrm{Macro}$). Superscripts and subscripts represent the $68\%$ credible interval.}
\label{tab:optimal_hyper_params}
\end{table*}
\section{Robustness Analysis of the Potential Correction Results for the Jackpot Lens}
\label{sec:appdx_B}
\begin{figure*}
	\includegraphics[width=\textwidth]{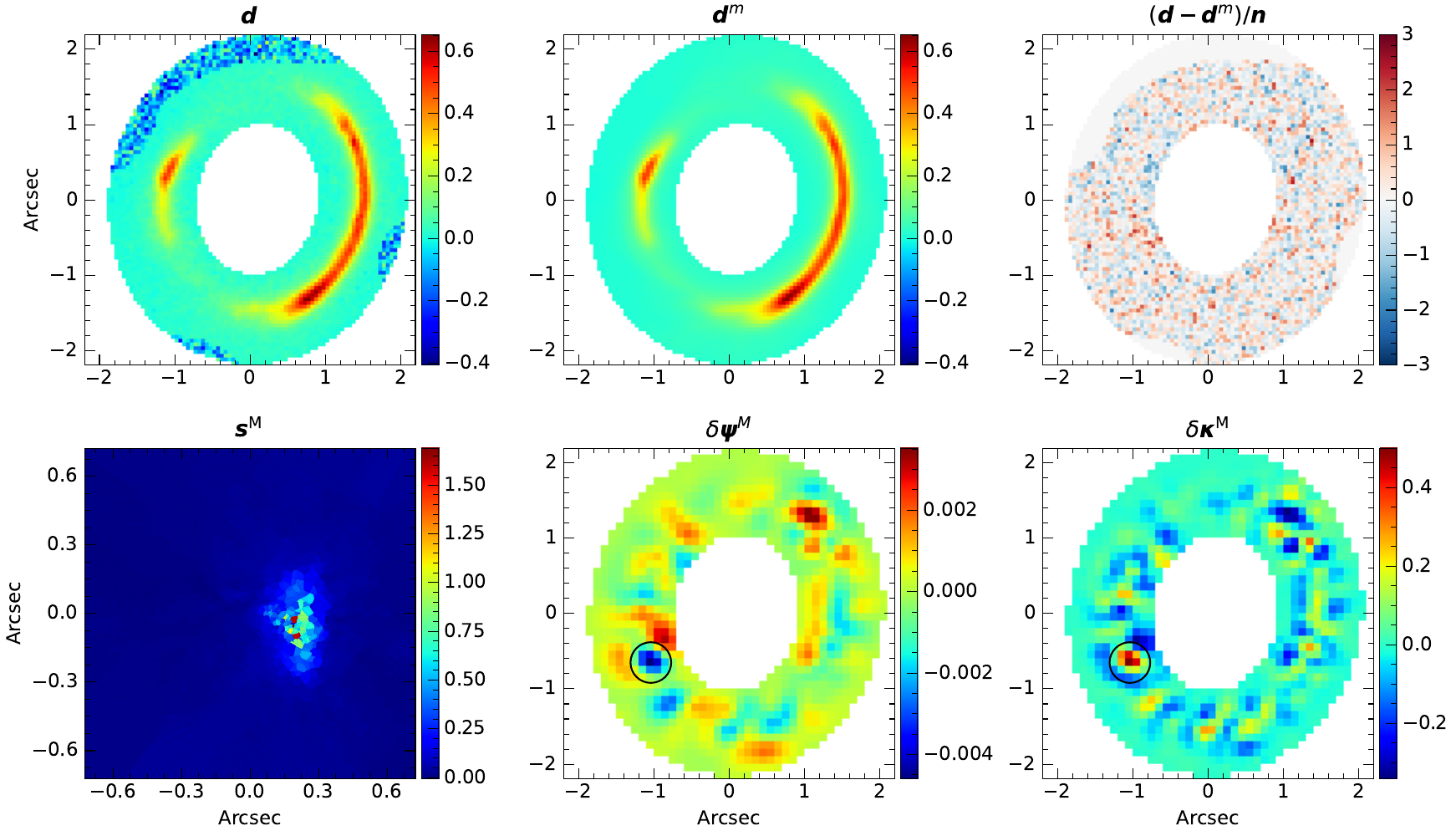}
        \caption{Similar to Figure~\ref{fig:dpsi_src_inv_0946}, but showing the potential correction result obtained with a larger annular mask.}
    \label{fig:0946_large_msk}
\end{figure*}

\begin{figure*}
	\includegraphics[width=\textwidth]{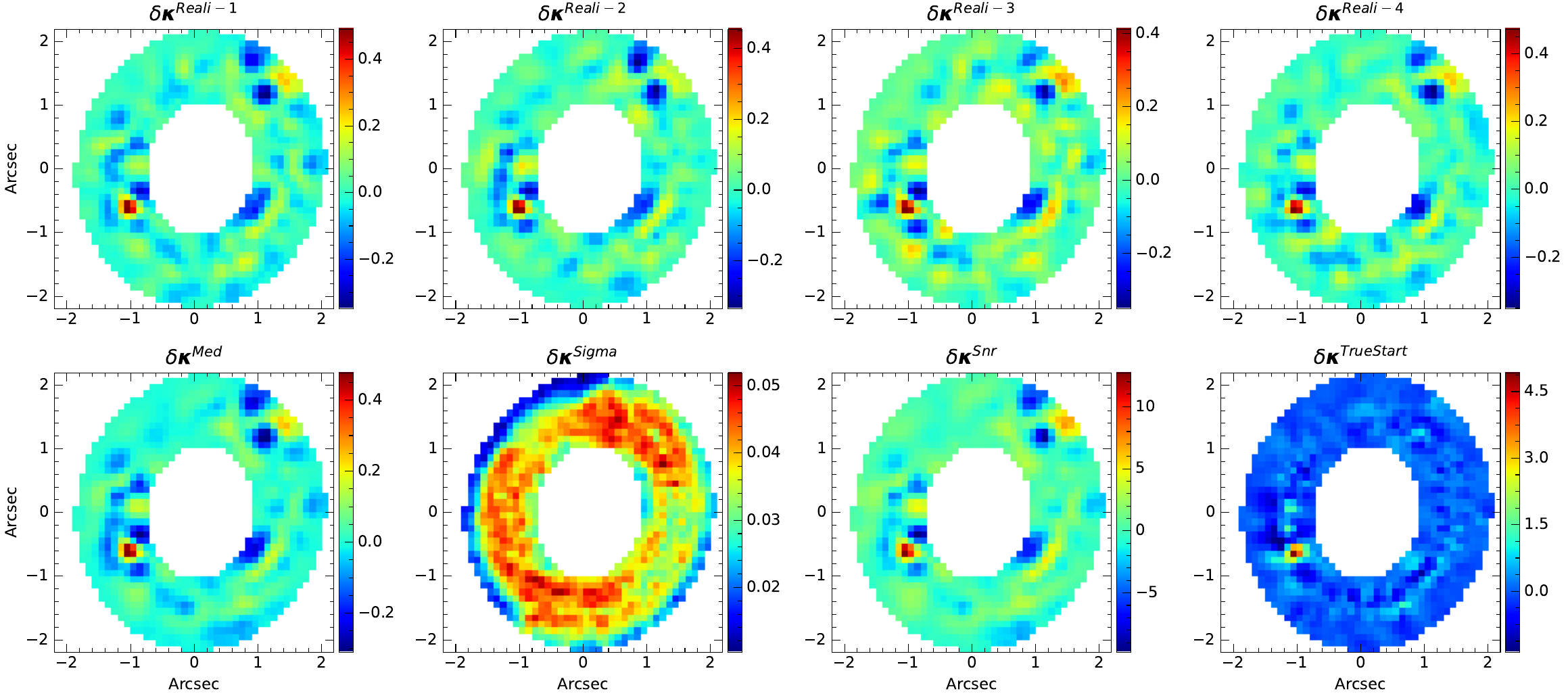}
        \caption{Assessment of perturbative convergence reconstruction using the potential correction method via Monte Carlo simulations. The top row displays convergence corrections for four mock Jackpot-like lenses differing only in noise realisation. The second row (left and middle-left panels) shows the mean and standard deviation of convergence corrections across 1000 mock Jackpot-like lenses with varying noise realisations. The signal-to-noise ratio (SNR) of the convergence correction, defined as the ratio of the mean to the standard deviation, quantifies the significance of the convergence correction. The bottom-right panel illustrates the convergence correction for a mock Jackpot-like lens initialised with the true main lens mass and source light distributions.}
    \label{fig:0946_mc_diagnosis}
\end{figure*}

\begin{figure*}
	\includegraphics[width=\textwidth]{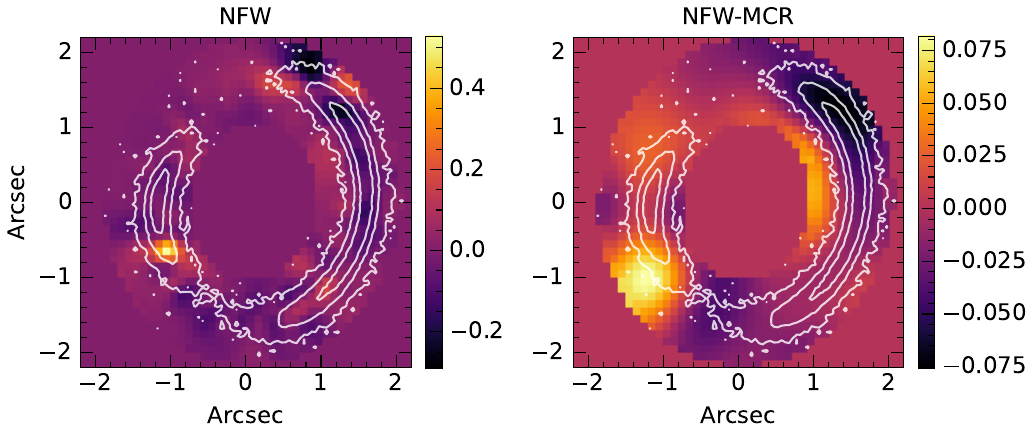}
        \caption{Convergence correction results for two Jackpot-like mock lenses, demonstrating the ability of our potential correction method to reveal the mass distribution of the input subhalos. The left lens includes an input NFW subhalo with a mass of $\sim 10^{10}\,M_{\odot}$ and a high concentration ($\sim 200$), centred at $(-1.040, -0.651)$ \citep{Despali2024}. The right lens contains an input NFW subhalo with a mass of $\sim 10^{11.5}\,M_{\odot}$, following the standard mass-concentration relation, centred at $(-1.00, -1.10)$ \citep{Nightingale2024}.}
    \label{fig:0946_compact_mock_check}
\end{figure*}

This section systematically evaluates the robustness of our potential correction method applied to the Jackpot lens. Figure~\ref{fig:0946_large_msk} presents the potential correction results obtained using a larger annular mask. Comparison of the resulting convergence perturbation map with that derived using a smaller mask (Figure~\ref{fig:dpsi_src_inv_0946}) shows that the properties of the reconstructed subhalo are consistent between the two cases. This demonstrates that our potential correction method is robust against the choice of mask size.

In addition to the localised mass clumps corresponding to the subhalo, additional mass features are observed in the bottom-right panel of Figures~\ref{fig:dpsi_src_inv_0946} and ~\ref{fig:0946_large_msk}, including a dipole-like structure near the detected subhalo and a negative convergence peak at $(x, y) \approx (1.0, 1.0)$. To investigate whether these additional mass features are genuine or induced by systematic errors in our potential correction method, we adopt a parametric approach analogous to that employed by \citet{Despali2024} in modelling the Jackpot lens. This approach incorporates an EPL plus shear model for the mass distribution of the main lens, an NFW model for the subhalo, and a pixelized source model. Subsequently, we simulate 1000 Jackpot-like mock lenses based on this model, each with a distinct realisation of image noise. Applying our potential correction method to these mock lenses consistently reproduces the additional mass features observed in the analysis of the real Jackpot lens data. We then utilise the median and standard deviation computed from the 1000 convergence perturbation maps (derived from the 1000 Jackpot-like mock lenses) as proxies for the signal and noise of the reconstructed mass perturbation features. The ratio between these quantities defines the signal-to-noise ratio (SNR), which quantifies the statistical significance of the reconstructed mass perturbation features. We find that the SNR of these additional mass features exceeds 3, confirming that they represent genuine artefacts induced by a systematic error in our potential correction method rather than being introduced by random noise (noting that these additional mass features should be absent in the mock lens data). If the potential correction is initiated using the true main lens mass model and source light model, the additional mass features vanish from the correction result, as illustrated in the bottom-right panel of Figure~\ref{fig:0946_mc_diagnosis}. We thus conclude that these additional features result from initiating the potential correction with a biased estimate of the main lens mass and source light provided by the macro model. Notably, the convergence correction shown in the bottom-right panel of Figure~\ref{fig:0946_mc_diagnosis} recovers the central high-density region of the cuspy subhalo more accurately than the correction initiated from a biased estimate (see Figure~\ref{fig:model_comparison_0946}). Hence, in addition to the limited dynamical range arising from the use of constant regularisation in the pixelized potential correction (discussed in Section~\ref{sec:discuss_adpt_reg}), initiating the correction with a biased main lens mass and source estimation also contributes to the underestimation of the central convergence of a cuspy subhalo.

A key strength of our potential correction method is its ability not only to detect the presence of a dark matter subhalo but also to reveal its spatial mass distribution. This capability allows us to conclude, using a model-independent approach, that the subhalo in the Jackpot lens is overconcentrated, potentially challenging the standard cold dark matter (CDM) model. To assess the robustness of our potential correction method for measuring subhalo compactness, we simulated two Jackpot-like mock lenses using subhalo properties from \citet{Despali2024} and \citet{Nightingale2024}, respectively, while keeping the main lens mass and source light distribution fixed. Applying our potential correction method to these mock lenses allowed us to detect a localised perturber in both cases (see Figure~\ref{fig:0946_compact_mock_check}). Moreover, the detected perturber was more compact in the mock lens based on the \citet{Despali2024} subhalo model. This demonstrates that our potential correction method can determine the compactness of the localised perturber.


\bsp	
\label{lastpage}
\end{document}